\documentclass[twocolumn,superscriptaddress]{revtex4-2}

\usepackage{bm}
\usepackage{graphicx}
\usepackage{amsmath}
\usepackage{hyperref}
\usepackage{color}

\begin{document}

\title{Nonequilibrium Green-Kubo relations for hydrodynamic transport from an equilibrium-like fluctuation-response equality}

\author{Hyun-Myung Chun}
\affiliation{Department of Biophysics, University of Michigan, Ann Arbor, Michigan, 48109, USA}

\author{Qi Gao}
\affiliation{Department of Physics, University of Michigan, Ann Arbor, Michigan 48109, USA}

\author{Jordan M. Horowitz}
\email{jmhorow@umich.edu}
\affiliation{Department of Biophysics, University of Michigan, Ann Arbor, Michigan, 48109, USA}
\affiliation{Department of Physics, University of Michigan, Ann Arbor, Michigan 48109, USA}
\affiliation{Center for the Study of Complex Systems, University of Michigan, Ann Arbor, Michigan 48104, USA}

\date{\today}

\begin{abstract}
Near equilibrium, Green-Kubo relations provide microscopic expressions for macroscopic transport coefficients in terms of equilibrium correlation functions.  At their core, they are based on the intimate relationship between response and fluctuations as embodied by the equilibrium fluctuation-dissipation theorem, a connection generically broken far from equilibrium. 
In this work, we identify a class of perturbations whose response around far-from-equilibrium steady states is linked to steady-state correlation functions via an equilibrium-like fluctuation-response equality.
We then utilize this prediction to substantiate linearized hydrodynamic transport equations that describe how spatial inhomogeneities in macroscopic nonequilibrium systems relax.
As a consequence, we derive nonequilibrium Green-Kubo relations for the transport coefficients of two types of hydrodynamic variables: local conserved densities and broken-symmetry modes. 
A byproduct of this work is to provide a theoretical foundation for the validity of Onsager's regression hypothesis around nonequilibrium steady states. 
Our predictions are analytically and numerically corroborated for two model systems: density diffusion in a fluid of soft, spherical active Brownian particles and phase diffusion in the noisy Kuramoto model on a square lattice.
\end{abstract}
\maketitle


\section{Introduction}
The fluctuation-dissipation theorem (FDT) is a cornerstone of statistical mechanics~\cite{callen1951irreversibility,kubo1966fluctuation,marconi2008fluctuation,forster1975hydrodynamic}.
It quantifies a fundamental correspondence between two experimental procedures for interrogating an equilibrium system: measuring a system's response to a weak perturbation and passively observing its fluctuations both contain the same information.
A major consequence has been in refining our understanding of the material coefficients that determine how spatial inhomogeneities in near-equilibrium macroscopic systems relax via hydrodynamic transport.
The resulting predictions, known as Green-Kubo relations, equate these macroscopic transport coefficients $D$ to the microscopic equilibrium correlation functions of local current observables $j_{\bm{r}}(t)$, whose fluctuations depend on space ${\bm r}$ and time $t$ in a volume $V$~\cite{green1954markoff,kubo1957statistical,kubo1957statistical2,forster1975hydrodynamic,kadanoff1963hydrodynamic,zwanzig1965timecorrelation,spohn1991large}:
\begin{equation}\label{eq:typical_GK}
    D\chi = \frac{\beta}{V} \int_0^\infty dt \int_V d\bm{r} \int_V d\bm{r}' ~ \langle j_{\bm{r}}(t) j_{\bm{r}'}(0) \rangle_{\rm eq}.
\end{equation}
Here, $\beta$ is the inverse temperature and $\chi$ is the static susceptibility (or thermodynamic derivative).

Even for far-from-equilibrium steady states, response can still be related to a nonequilibrium correlation function by a modification of the FDT's original derivation~\cite{agarwal1972fluctuation,seifert2010fluctuation}.
The resulting correlation function turns out to be rather formal, often requiring detailed microscopic knowledge of the steady state or its dynamics~\cite{speck2006restoring,baiesi2009fluctuations,prost2009generalized,chetrite2009eulerian,altaner2016fluctuation}.
Even still, this nonequilibrium modification of the FDT allows one to link integrals of nonequilibrium correlation functions to microscopic currents~\cite{seifert2010generalized,asheichyk2019response,dal2019linear}, in much the same spirit as the macroscopic Green-Kubo relation in~\eqref{eq:typical_GK}.

Nonequilibrium Green-Kubo relations that relate macroscopic transport coefficients to local current observables have appeared in the literature in particular situations.  
These studies can be categorized according to the method.  
For a two-dimensional nonequilibrium viscous fluid~\cite{epstein2020time,hargus2020time}, Green-Kubo relations were deduced by assuming that Onsager's regression hypothesis~\cite{onsager1931reciprocal} remains valid around nonequilibrium steady states.  An alternative approach utilizes the projection operator method~\cite{mori1965transport,zwanzig2001nonequilibrium} adapted for non-Hamiltonian dynamics~\cite{ernst2005generalized,ernst2006new,espanol2002coarse,espanol2009einstein}.
The resulting Green-Kubo relations incorporate a time-reversed dynamics, apparently obscuring the interpretation of  the resulting correlation functions.
This obstacle has been overcome for at least one specific model of a nonequilibrium active fluid~\cite{han2020statistical}.

In this work, we demonstrate that generally Green-Kubo relations for macroscopic transport coefficients maintain their equilibrium form arbitrarily far from equilibrium.
The key step in this derivation is the elucidation of a class of perturbations with explicit conjugate variables whose response is given by simple nonequilibrium correlation functions, akin to the equilibrium FDT.
This observation generalizes our previous work on static response to time-dependent perturbations~\cite{owen2020universal}.
We then exploit this equilibrium-like fluctuation-response equality to provide a theoretical foundation for linearized hydrodynamic equations governing transport in homogenous nonequilibrium fluids.
The technique we use was originally developed by Kubo to analyze conductivity~\cite{kubo1957statistical}, but elaborated for simple equilibrium fluids by Oppenheim and collaborators~\cite{felderhof1965correlation,selwyn1971generalized,weare1974nonlinear,oppenheim1971linear}.
We consider two kinds of slow hydrodynamic modes, local densities of conserved variables and the Nambu-Goldstone modes that emerge from a broken continuous symmetry~\cite{forster1975hydrodynamic}.
Our theory is illustrated by numerical simulations of two examples, a fluid of soft active Brownian particles and a noisy Kuramoto model.

\section{Fluctuation-response equality}\label{sec:FRE}
\begin{figure}[t]
\centering
\includegraphics[width=\columnwidth]{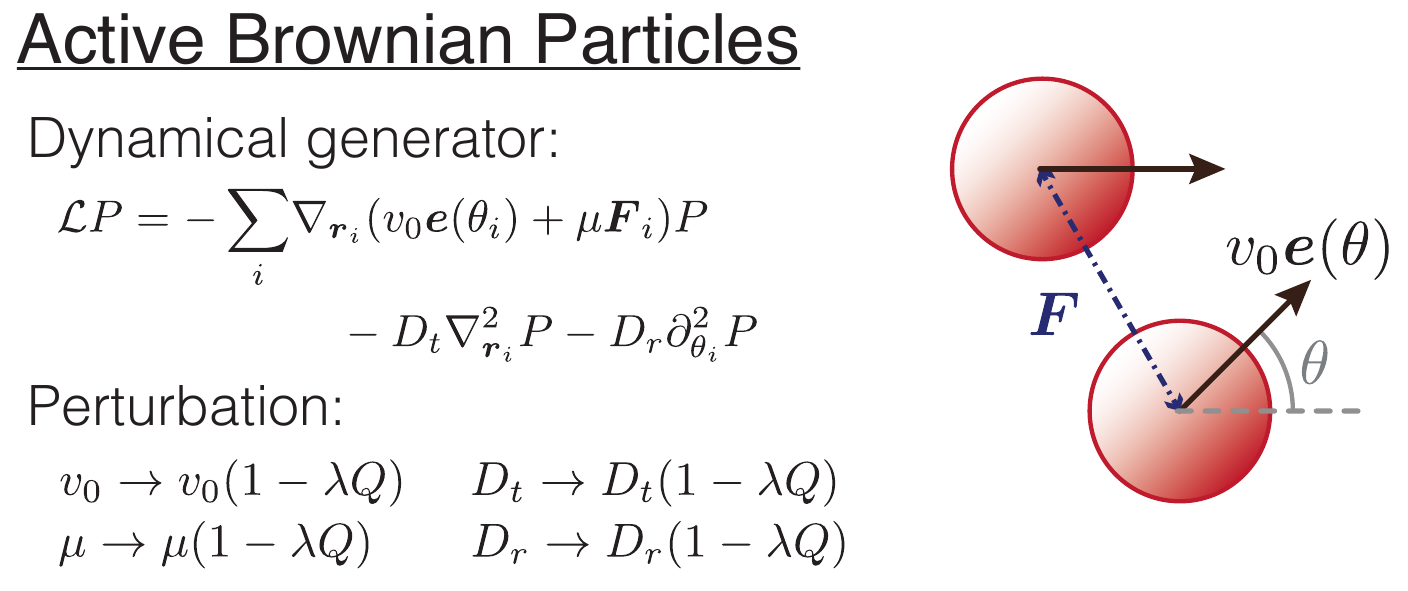}
\caption{Perturbing the dynamics of a fluid of active Brownian particles: Particles (red) are self-propelled with a velocity $v_0$ whose direction $\theta$ diffuses.  They interact pair-wise through a repulsive force $F$ and experience translational noise. The surrounding environment induces diffusive behavior of the particles characterized by diffusion coefficients $D_t$ and $D_r$. The stochastic equations of motion~\eqref{eq:ABP} correspond to the Fokker-Planck operator ${\mathcal L}$.  The type of perturbation in~\eqref{eq:perturbed_equation} corresponds to multiplying all system parameters by $1-\lambda Q$.}
\label{fig:ABP}
\end{figure}

Consider a classical system with microscopic state $\bm{z}$, whose dynamics, either stochastic or deterministic, are well-modeled as Markovian.
As such, the system's probability density $P(\bm{z},t)$ evolves according to the Master equation~\cite{gardiner2009stochastic}
\begin{equation}\label{eq:master_eq}
    \partial_t P(\bm{z},t) = {\mathcal{L}} P(\bm{z},t),
\end{equation}
with generator ${\mathcal L}$. 
We will further assume that left undisturbed the system will relax to a unique steady-state distribution  $P_{\rm ss}({\bm z})$ given by the solution of ${\mathcal L}P_{\rm ss}=0$.
This model incorporates deterministic Hamiltonian dynamics, where ${\mathcal L}P=\{P,H\}$ is the Poisson bracket with a Hamiltonian $H$, as well as non-Hamiltonian deterministic and stochastic dynamics.
A stochastic nonequilibrium example that will be a useful illustration is the fluid of active Brownian particles (ABPs) pictured in Fig.~\ref{fig:ABP}, where ${\mathcal L}$ is the Fokker-Planck operator~\cite{gardiner2009stochastic}.

Our focus is on how averages of observables $\langle O(t)\rangle = \int  O({\bm z}) P({\bm z},t)d{\bm z}$ change in response to weak perturbations.
While this can be analyzed quite generally~\cite{agarwal1972fluctuation,baiesi2009fluctuations,prost2009generalized,seifert2010fluctuation}, we have identified a specific class of dynamical perturbations with interesting and useful properties of the form
\begin{equation}\label{eq:perturbed_equation}
 {\mathcal L}'P={\mathcal L}P -
    \lambda(t)\mathcal{L} (Q({\bm z}) P),
\end{equation}
with a small time-dependent parameter $\lambda(t)$ and a time-independent conjugate coordinate $Q({\bm z})$.
For an equilibrium system evolving under Hamiltonian dynamics, the perturbation takes the form $\lambda {\mathcal L}(QP)=\lambda\{QP,H \}$, which at first glance appears to differ from the classical linear response perturbation $\lambda \{P,Q\}$ generated by modifying the Hamiltonian $H'=H - \lambda Q$.
However, at equilibrium, the distribution is solely a function of the Hamiltonian, like for example the canonical distribution $P_{\rm eq}\propto e^{-H}$ (with $\beta = 1$).
In this case, both perturbations have the same effect on an equilibrium distribution, which can be seen by noting that $\{Q P_{\rm eq},H\}= \{Q,H\} P_{\rm eq} = \{ P_{\rm eq}, Q \}$, since $\{P_{\rm eq},H\}=0$.
Thus, this approach encompasses equilibrium linear response theory.
Clearly, multiple perturbations can then lead to the same response, and in Appendix~\ref{sec:equivalence}, we delineate in detail this equivalence class.
For nonequilibrium dynamics, implementing the perturbation embodied in~\eqref{eq:perturbed_equation} can require a rather complicated coordinated change in the system's parameters, as can be seen in Fig.~\ref{fig:ABP}.

Now, imagine starting the system in its nonequilibrium steady state at $t=-\infty$ and then turning on the perturbation.
A standard first-order perturbation analysis~\cite{agarwal1972fluctuation,seifert2010fluctuation,risken2012fokker} detailed in Appendix~\ref{sec:equivalence} reveals that the average of an observable $\langle O(t)\rangle$ at time $t$ over a statistical ensemble initially at steady state will deviate from the steady-state average $\langle O\rangle_{\rm ss} = \int  O({\bm z}) P_{\rm ss}({\bm z})d{\bm z}$ by
\begin{equation}\label{eq:generalized_FDT}
    \langle O(t) \rangle - \langle O \rangle_{\rm ss}
    = \int_{-\infty}^t 
    \left\langle O(t) {\dot Q}(s) \right\rangle_{\rm ss}  \lambda(s)\ ds .
\end{equation}
Here, we have introduced the time-translationally-invariant steady-state correlation function, defined for any two observables, $O_1(\bm{z})$ and $O_2(\bm{z})$,  by
\begin{equation}\label{eq:corr}
\begin{aligned}
    & \langle O_1(t) O_2(s) \rangle_{\rm ss} \\
    & \qquad= \int d{\bm z}d{\bm z}'\ O_1({\bm z})P({\bm z}, t|{\bm z}',s) O_2({\bm z}')P_{\rm ss}({\bm z}') \\
    & \qquad=\int d{\bm z}\ O_1({\bm z})e^{(t-s){\mathcal L}} \left(O_2({\bm z})P_{\rm ss}({\bm z})\right)
\end{aligned}
\end{equation}
in terms of the transition probability $P({\bm z}, t|{\bm z}',s)$ obtained as the solution of \eqref{eq:master_eq} with a delta-function initial condition $\delta({\bm z}-{\bm z}')$.

Noteworthy is that the response to these perturbations is given by the steady-state correlation between the observable $O$ and the conjugate observable $Q$ that explicitly appears in the perturbation~\eqref{eq:perturbed_equation}, just like the equilibrium FDT.
By contrast, previous studies that addressed generic perturbations, arrived at correlation functions where instead of $\dot{Q}$, the conjugate observable that appeared required knowledge of the entire steady-state distribution (via $\partial_\lambda\ln P_{\rm ss}$) or of the generator of the dynamics~\cite{agarwal1972fluctuation,baiesi2009fluctuations,prost2009generalized,seifert2010fluctuation}.
 
While executing such a perturbation in the lab will generically be challenging, the theoretical utility is born out in the simplicity of the response function and the applications that entails.
Firstly, the fluctuation-response equality in~\eqref{eq:generalized_FDT} implies we can measure any correlation function we choose by observing the appropriate response~\eqref{eq:perturbed_equation}.
We can exploit this in a computational experiment, where there is no issue applying any perturbation we choose. 
Such an approach would be the far-from-equilibrium equivalent of the ``nonequilibrium perturbation method'' utilized for measuring equilibrium correlation functions in simulation~\cite{allen1989computer}.
The second major application, and the one we focus on for the rest of this paper, is to derive Green-Kubo relations. 
Some of the earliest derivations of Green-Kubo relations exploited the equilibrium FDT \cite{kadanoff1963hydrodynamic,felderhof1965correlation,zwanzig1965timecorrelation}.
It was very quickly realized that the external perturbations played only a formal intermediary role in the derivation and therefore need not be realizable in the lab~\cite{felderhof1965correlation}.
A concrete example of this point is the derivation of the Green-Kubo relation for the shear viscosity of a simple fluid due to Jackson and Mazur where a non-gradient force was introduced into a Hamiltonian dynamics to mimic a shear~\cite{jackson1964statistical}.
With this observation, we now employ this previously developed program to study transport in macroscopic systems, adapting it to nonequilibrium steady states using our fluctuation-response equality~\eqref{eq:generalized_FDT}.

\section{Hydrodynamic transport}
We now focus our attention on an $N$-particle macroscopic system in a volume $V$.
We will assume that its macroscopic dynamics on long length- and time-scales can be accurately captured by a few hydrodynamic variables, which are many-bodied observables with a relaxation time that diverges with system size.
Typically they are local densities of conserved quantities or Nambu-Goldstone modes due to broken continuous symmetries~\cite{forster1975hydrodynamic}.
For example, the near-equilibrium dynamics of a simple fluid are completely captured by the conserved number, momentum, and energy densities~\cite{forster1975hydrodynamic}. 
Away from equilibrium, hydrodynamics has been quite successful in modeling active matter~\cite{marchetti2013active}---systems composed of constituents that individually consume energy.
Theories of flocking~\cite{toner1998flocks}, active gels~\cite{prost2015active}, and experiments on cellular spindle dynamics~\cite{brugue2014spindle} further reinforce the need to describe both conserved densities as well as broken-symmetry modes.
For clarity of presentation, we will focus on systems that possess a single hydrodynamic variable $A_{\bm r}(t)$; the generalization to multiple hydrodynamic variables is presented in Appendix~\ref{sec:gen_theory}.

In the homogeneous steady state, our hydrodynamic variable will be spatially uniform and constant in time with value $\bar{A}$.
Upon the emergence of a small spatial inhomogeneity, either due to an external stimulus or internal fluctuation, the deviation from the steady-state $\delta A_{\bm r}(t)=A_{\bm r}(t)-{\bar A}$ will by assumption relax (or regress) via a linear hydrodynamic equation with drift ${\bm v}$ and transport coefficient matrix ${\mathsf D}$,
\begin{equation}\label{eq:hydroEq}
\partial_t {\delta A}_{\bm r}(t) = -\nabla\cdot \left({\bm v}\ \delta A_{\bm r}(t)\right)+\nabla \cdot {\mathsf D}\cdot \nabla \delta A_{\bm r}(t),
\end{equation}
or in Fourier space ($\delta A_{\bm k}(t) = \int_V \delta A_{\bm r}(t) e^{i{\bm k}\cdot{\bm r}}\ d{\bm r}$), 
\begin{equation}\label{eq:hydroEqFourier}
\partial_t{ \delta A}_{\bm k}(t) = \left(i{\bm k}\cdot{\bm v} - {\bm k}\cdot {\mathsf D}\cdot {\bm k}\right) \delta A_{\bm k}(t).
\end{equation}
Near equilibrium when there is a single hydrodynamic variable ${\bm v}=0$, due to time-reversal symmetry, as there can be no preferred direction of motion.
To have a nontrivial drift ${\bm v}\neq 0$ near equilibrium, multiple hydrodynamics are required, in which case ${\bm v}$ is called the Eulerian term or the reactive coupling~\cite{chaikin1995principles}.
Nonequilibrium steady states, by contrast, can support spatial flow of a single variable and therefore we allow for a nonzero ${\bm v}$ here.

Equation~\eqref{eq:hydroEq} (or equivalently \eqref{eq:hydroEqFourier}) serves as the definition of the parameters ${\bm v}$ and ${\mathsf D}$. 
As such, at the macroscopic hydrodynamic level, they can only be measured by first setting up a spatially inhomogeneous profile and fitting the subsequent relaxation to Eq.~\eqref{eq:hydroEq} (or \eqref{eq:hydroEqFourier}).

We can now view~\eqref{eq:hydroEqFourier} as the beginning of a long wavelength (small $k=|{\bm k}|$) expansion of a generalized linear transport equation modeling the exponential regression of the hydrodynamic variable:
\begin{equation}\label{eq:generalTransport}
\partial_t {\delta A}_{\bm k}(t) =-M_{\bm k} \delta A_{\bm k}(t),
\end{equation}
where $M_{\bm k}$ is a generalized transport coefficient that for small-$k$ must behave as $M_{\bm k} \approx -i{\bm k}\cdot{\bm v}+{\bm k}\cdot {\mathsf D}\cdot {\bm k}+\cdots$.
In the following section, we will use our equilibrium-like fluctuation-response relation to provide a statistical-mechanical basis for this expansion and as a result extract microscopic expressions for ${\bm v}$ and ${\mathsf D}$, known as Green-Kubo relations.

But first we need to connect the macroscopic to the microscopic.
To this end, we specify our hydrodynamic system's microscopic state.
For each of the $i=1,\dots, N$ particles, we denote its position as ${\bm r}_i$ and any other of its degrees of freedom, like momentum or polarity, as $\bm{s}_i$, such that  ${\bm z}_i = ({\bm r}_i,{\bm s}_i)$ and ${\bm z} = (\bm{z}_1,\cdots,\bm{z}_N)$.
Microscopic expressions for hydrodynamic variables are then typically formed as densities of single-particle observables $a_j({\bm z}_j)$ through a sum of the form~\cite{forster1975hydrodynamic}
\begin{equation}
    {\hat A}_{\bm{r}}(\bm{z}) = \sum_{j=1}^N a_j({\bm z}_j) \delta(\bm{r} - \bm{r}_j).
\end{equation}
For example, the choice $a_j = 1$ defines the particle number density.
Notice that the $\{{\hat A}_{\bm r}({\bm z})\}_{{\bm r}\in V}$ are a family of state-space observables parameterized by the spatial location ${\bm r}$.
The link to the macroscopic hydrodynamic variable then emerges on average, $\langle {\hat A}_{\bm r}(t)\rangle =A_{\bm r}(t)$. 

The two types of hydrodynamic variables we will consider---conserved densities and Nambu-Goldstone modes---are identified by their microscopic dynamics.
For the conserved density, its evolution along a single dynamical trajectory is distinguished by a continuity equation $\partial_t {\hat A}_{\bm{r}} + \bm{\nabla}_{\bm{r}} \cdot \bm{j}_{\bm{r}} = 0$, with corresponding local current $\bm{j}_{\bm r}$.
Re-expressing this continuity equation in Fourier space, $\partial_t {\hat A}_{\bm{k}} = i\bm{k} \cdot \bm{j}_{\bm{k}}$, is particularly illuminating, since it makes explicit that the small-$k$ modes vary slowly in time with a relaxation time that diverges as $k\to 0$.
By contrast, Nambu-Goldstone modes need not be conserved.  
Instead, they are distinguished by diverging static fluctuations.

\section{Green-Kubo relations}
The generalized transport equation in~\eqref{eq:generalTransport} describes the relaxation from an initially inhomogeneous profile.
To connect this relaxation to the microscopic dynamics, we follow the procedure laid out in~\cite{felderhof1965correlation,selwyn1971generalized,weare1974nonlinear}. 
Underpinning this procedure is the conceptual insight that if we start to slowly turn on an external perturbation conjugate to $A_{\bm r}$  from $t=-\infty$, by $t=0$ we have generated an inhomogeneous profile that is minimally disturbed from the steady state.
At $t=0$, we switch off the perturbation and track the subsequent evolution.
If this microscopic experiment is to be consistent with the macroscopic transport equation, the two must both give the same evolution on long length- and time-scales.
This consistency requirement leads to nonequilibrium Green-Kubo relations connecting the relaxation dynamics to fluctuations.

Our first step is to specify the coordinate in our perturbation~\eqref{eq:perturbed_equation}.
The required form is 
 \begin{equation}\label{eq:Q}
 Q({\bm z}) = \int _V {\hat A}_{\bm r}({\bm z}) f_{\bm r}\ d{\bm r},\qquad \lambda(t) = e^{\epsilon t}\Theta(-t),
 \end{equation}
where we eventually take $\epsilon\to 0^+$, and the Heaviside step function $\Theta(-t)$ turns off the perturbation at $t=0$. 
The choice of $f_{\bm r}$ is immaterial as long as all integrals converge, as it will shortly drop out of the calculation.
Using this choice of conjugate coordinate in the fluctuation-response equality~\eqref{eq:generalized_FDT} leads to an expression for the evolution of ${\hat A}$ for $t \ge 0$, which after an integration by parts and sending $\epsilon\to 0$, becomes
\begin{equation}\label{eq:GKStep1}
\begin{aligned}
    \langle \delta {\hat A}_{\bm{r}}(t) \rangle
    =  \int_V  
    \langle  {\hat A}_{\bm{r}}(t) {\hat A}_{\bm{r}'}(0) \rangle_{\rm ss}
    f_{\bm{r}'}\ d\bm{r}'.
\end{aligned}
\end{equation}
We now remove the dependence on $f_{\bm r}$ in favor of the nonuniform profile generated by this perturbation at $t=0$.
This is facilitated by first Fourier transforming~\eqref{eq:GKStep1} and exploiting the assumed translational invariance of the homogenous steady state to arrive at 
\begin{equation}\label{eq:EQ_FDT}
\begin{aligned}
    \langle \delta {\hat A}_{\bm{k}}(t) \rangle
    = \frac{1}{V} \langle {\hat A}_{\bm{k}}(t) {\hat A}_{-\bm{k}}(0) \rangle_{\rm ss}
    f_{\bm{k}}
\end{aligned}
\end{equation}
in the large system-size limit~\cite{felderhof1965correlation}.
This equality is true at every positive time, including at $t=0$, which allows us to solve for $f_{\bm{k}}=V\langle \delta {\hat A}_{\bm k}(0)\rangle / \langle {\hat A}_{\bm k}(0){\hat A}_{-{\bm k}}(0)\rangle_{\rm ss}$ in terms of the initial value $\langle \delta \hat{A}_{\bm{k}}(0) \rangle$ and substitute back in to find
\begin{equation}\label{eq:Onsager_hypothesis}
\begin{aligned}
   \frac{ \langle \delta {\hat A}_{\bm{k}}(t) \rangle}{\langle \delta {\hat A}_{\bm{k}}(0) \rangle}
    = \frac{\langle {\hat A}_{\bm{k}}(t) {\hat A}_{-\bm{k}}(0) \rangle_{\rm ss}}
    {\langle {\hat A}_{\bm{k}}(0) {\hat A}_{-\bm{k}}(0) \rangle_{\rm ss}}. 
\end{aligned}
\end{equation}
Equality~\eqref{eq:Onsager_hypothesis} demonstrates that the time evolution of the relaxation of the average is identical to that of the correlation function, at least for the particular ensemble produced at $t=0$ by this perturbation.

Importantly, equality~\eqref{eq:Onsager_hypothesis} depends only on the behavior of the sole slow hydrodynamic variable.
(Here, there is just one variable, but in general we need to include all slow degrees of freedom as described in Appendix~\ref{sec:gen_theory}.)
Thus, at long enough length- and time-scales, any vagaries of the initial preparation will die out exponentially fast, and~\eqref{eq:Onsager_hypothesis} should continue to remain true when the generalized transport equation in~\eqref{eq:hydroEq} accurately describes the exponential relaxation of $A_{\bm r}(t)=\langle {\hat A}_{\bm r}(t)\rangle$.
From this we can conclude that the steady-state correlation function of the hydrodynamic mode is governed by the same macroscopic linear equation,
\begin{equation}\label{eq:Onsager_hypothesis2}
    \partial_t \langle {\hat A}_{\bm{k}}(t) {\hat A}_{-\bm{k}}(0) \rangle_{\rm ss}
    = -M_{\bm{k}} \langle {\hat A}_{\bm{k}}(t) {\hat A}_{-\bm{k}}(0) \rangle_{\rm ss},
\end{equation}
at least for long times and small $k$.
This prediction is a key contribution of this study as it provides a statistical-mechanical rationale for an Onsager's regression hypothesis around nonequilibrium steady states, as was conjectured in~\cite{hargus2020time,epstein2020time}.

Consistency between the macroscopic and microscopic descriptions embodied by~\eqref{eq:Onsager_hypothesis2} leads to the conclusion that $M_{\bm k}$ can be inferred from the dynamics of steady-state correlation functions~\cite{felderhof1965correlation,selwyn1971generalized,weare1974nonlinear}.
Thus, we can exploit~\eqref{eq:Onsager_hypothesis2} to solve for the generalized transport coefficient systematically for small $k$, when we expect hydrodynamics to be an accurate description, and thus deduce the expansion parameters $M_{\bm k} \approx -i{\bm k}\cdot{\bm v}+{\bm k}\cdot {\mathsf D}\cdot {\bm k}+\cdots$.
The steps are detailed in Appendix~\ref{sec:gen_theory}, but follow closely~\cite{felderhof1965correlation,selwyn1971generalized,weare1974nonlinear}.
Here, we report the results.

The precise form of the resulting Green-Kubo relations depends on the specifics of how the hydrodynamic variable's microscopic dynamics behave for small wave numbers.
Let us first address conserved local densities whose microscopic relaxation rate diverges with small $k$, $\partial_t{\hat A}_{\bm k}=i{\bm k}\cdot{\bm j}_{\bm k}$,  but whose static correlation function remains finite, $\tilde\chi=\lim_{{\bm k\to 0}}\langle \hat{A}_{\bm k} \hat{A}_{-{\bm k}}\rangle_{\rm ss}$.
The static correlation function $\tilde{\chi}$ is related to the static structure factor by dividing by an extensive quantity such as $V$ or $N$.
In equilibrium, $\tilde\chi/V$ can be related to a thermodynamic susceptibility via the equilibrium FDT, which can then be deduced solely from the equilibrium equation of state~\cite{kadanoff1963hydrodynamic,forster1975hydrodynamic}.
Equality~\eqref{eq:EQ_FDT} at $t = 0$ shows that $\tilde\chi$ can still be connected to the static response induced by~\eqref{eq:perturbed_equation}.
However, such an interpretation loses much of its utility away from equilibrium due to the absence of any macroscopic nonequilibrium thermodynamics.
Even still, recent work suggests there might be a useful thermodynamic structure, at least for spherical ABPs~\cite{dulaney2021isothermal}.

Carrying out the expansion of the generalized transport coefficient, we find for the drift
\begin{equation}\label{eq:GK_cons_v}
\bm{v} \tilde\chi = \lim_{k\to0} \langle \bm{j}_{\bm{k}} {\hat A}_{-\bm{k}} \rangle_{\rm ss},
\end{equation}
which captures the conservative (or Eulerian) part of the dynamics.
The transport coefficient then captures the fluctuations around this steady drift, which motivates the introduction of the dissipative current as the relative flow: ${\bm I}_{\bm r} = {\bm j}_{\bm r}-{\bm v}{\hat A}_{\bm r}$.
The dissipative current ${\bm I}_{\bm r}$ represents the rapidly fluctuating part of ${\bm j}_{\bm r}$, in light of the orthogonality $\lim_{k\to0}\langle {\bm I}_{\bm k}{\hat A}_{-{\bm k}}\rangle_{\rm ss}=0$, suggesting that its fluctuations decay on microscopic time-scales~\cite{weare1974nonlinear}.
The resulting Green-Kubo relation is most simply expressed using the half-Fourier transform in time ${\bm I}_{{\bm k}\omega}=\int_0^\infty {\bm I}_{\bm k}(t)e^{i\omega t}dt$:
\begin{equation}
\begin{aligned}\label{eq:GK_cons}
  {\mathsf D}\tilde\chi={\mathsf C}+{\mathsf E},
\end{aligned}
\end{equation}
where the first term is the dissipative-current correlation function
\begin{equation}
\begin{split}
 {\mathsf C} &= \lim_{\omega\to 0}\lim_{k\to 0}\langle {\bm I}_{{\bm k}\omega} {\bm I}_{-{\bm k}}\rangle_{\rm ss}\\
 &= \int_0^\infty dt \int_V d\bm{r} \int_V d\bm{r}'\  \langle {\bm I}_{\bm{r}}(t){\bm I}_{\bm{r}'}(0) \rangle_{\rm ss},\\
 \end{split}
\end{equation}
and the second is the matrix of first-order corrections to the drift
\begin{equation}
\begin{aligned}
{\mathsf E}^{\mu\nu}
& = \lim_{k\to0} \frac{\langle I_{\bm{k}}^\mu {\hat A}_{-\bm{k}} \rangle_{\rm ss}}{ik^\nu} \\
& = \int_V d\bm{r} \int_V d\bm{r}' ~ \langle I_{\bm{r}}^\mu \hat{A}_{\bm{r}'} \rangle_{\rm ss} (r^\nu - r'^\nu).
\end{aligned}
\end{equation}
The dissipative-current correlation function is assumed to decay fast enough so that the time-integral converges to a finite value.
As such, we are assuming that the plateau value problem, a long-standing problem of Green-Kubo relations~\cite{kirkwood1946statistical,mori1965transport,espanol1993force,mazenko2008nonequilibrium,espanol2019solution}, does not occur in systems of interest.

Comparison with the typical equilibrium expression in~\eqref{eq:typical_GK} brings to light two apparent differences.
The first is the use of the dissipative current ${\bm I}_{\bm r}$ instead of the current ${\bm j}_{\bm r}$, and the second is the appearance of ${\mathsf E}$.
Both terms had already appeared in early works on near-equilibrium transport~\cite{jackson1964statistical,selwyn1971generalized,weare1974nonlinear}, and were rediscovered for hard-core and Langevin dynamics~\cite{ernst2005generalized,ernst2006new,jung2016computing}, an observation we recount in more detail in Sec.~\ref{sec:earlier}.
Historically ${\mathsf E}$ was set to zero since early studies focused on near-equilibrium systems evolving with deterministic Hamiltonian dynamics where time-reversal symmetry forces $\langle {\bm j}_{\bm k} {\hat A}_{-{\bm k}}\rangle_{\rm ss}=0$~\cite{kubo1957statistical,jackson1964statistical}.
However, if the dynamics are stochastic it is possible for ${\mathsf E}\neq 0$ even in time-reversal-symmetric equilibrium systems, as we will see for an equilibrium fluid of Brownian particles.

Nambu-Goldstone modes, by contrast, are identified by the divergence of their static correlation function $\langle {\hat A}_{\bm k} {\hat A}_{-{\bm k}}\rangle_{\rm ss}$ as $k\to 0$.
Near equilibrium, rigorous arguments based on Goldstone's theorem and the Bogoliubov-inequality lead to a minimum divergence of $1/k^2$~\cite{forster1975hydrodynamic}.
However, for stochastic nonequilibrium dynamics, the microscopic foundation of these modes is shakier.
The connection between symmetries and conservation laws arising from Noether's theorem is not robust enough for stochastic dynamics to encompass typical nonequilibrium matter~\cite{baes2013noether}.
Without that connection there are no general grounds for a nonequilibrium Goldstone theorem.
Progress has been made on specific models using a field-theoretic approach~\cite{minami2018spontaneous}, whose general applicability is unclear to us.
Even still our approach allows us to demonstrate that diverging static fluctuations will generically lead to hydrodynamic modes in a nonequilibrium system.

To be as agnostic as possible, we take a generic divergence of the form $\langle \hat{A}_{\bm{k}} \hat{A}_{-\bm{k}} \rangle_{\rm ss} \sim k^{-q}$ with $q$ arbitrary.
To have a consistent small $k$ expansion of~\eqref{eq:Onsager_hypothesis2} when the static correlation function diverges, the correlation functions $\langle \dot{\hat{A}}_{\bm{k}\omega} \dot{\hat{A}}_{-\bm{k}}\rangle_{\rm ss}$ and $\langle \dot{\hat{A}}_{\bm{k}} \hat{A}_{-\bm{k}} \rangle_{\rm ss}$ must scale in such a way that the limits defining the drift and the leading order transport coefficient are finite:
\begin{equation}\label{eq:GK_NG_modes_v}
    {\hat {\bm k}}\cdot{\bm v} = \lim_{\omega\to 0}\lim_{k\to 0}
    \frac{\langle \dot{\hat{A}}_{\bm{k}} \hat{A}_{-\bm{k}} \rangle_{\rm ss} - \langle \dot{\hat{A}}_{\bm{k}\omega} \dot{\hat{A}}_{-\bm{k}} \rangle_{\rm ss}}
    {ik \langle \hat{A}_{\bm{k}} \hat{A}_{-\bm{k}} \rangle_{\rm ss}},
\end{equation}
\begin{equation}\label{eq:GK_NG_modes}
  {\hat{\bm k}}\cdot  {\mathsf D}\cdot{\hat{\bm k}}
    = \lim_{\omega\to 0} \lim_{k\to0}
    \frac{\langle \dot{\hat{A}}_{\bm{k}\omega}' \dot{\hat{A}}_{-\bm{k}}\rangle_{\rm ss}-\langle \dot{\hat{A}}_{\bm{k}}' \hat{A}_{-\bm{k}} \rangle_{\rm ss}}
    {k^2 \langle \hat{A}_{\bm{k}} \hat{A}_{-\bm{k}} \rangle_{\rm ss}}
\end{equation}
with $\hat{\bm{k}} = \bm{k}/k$ and $\dot{\hat{A}}_{\bm{k}}'(t) = \dot{\hat{A}}(t) - i({\bm k}\cdot{\bm v}) \hat{A}_{\bm{k}}(t)$.
These expressions are actually the most general forms for ${\bm v}$ and $\mathsf{D}$, and specialize to~\eqref{eq:GK_cons_v} and~\eqref{eq:GK_cons} when the hydrodynamic mode is conserved.

\section{Illustrations}
The Green-Kubo relations in \eqref{eq:GK_cons_v}, \eqref{eq:GK_cons}, \eqref{eq:GK_NG_modes_v}, and \eqref{eq:GK_NG_modes} predict that two distinct experimental procedures must give the same result: (i) measure the transport parameters ${\bm v}$ and ${\mathsf D}$ directly via their defining equation \eqref{eq:hydroEq} by perturbing the system and then measuring the rate of exponential relaxation, or (ii) extract the same information by passively observing the steady-state fluctuations.
In this section, we corroborate and illustrate this equivalence.
Our first model is a fluid of ABPs with a single conserved density, and the second is a noisy Kuramoto model, where the breaking of the continuous rotational symmetry upon synchronization leads to a Nambu-Goldstone mode.

\begin{figure*}[t]
	\includegraphics[width=2\columnwidth]{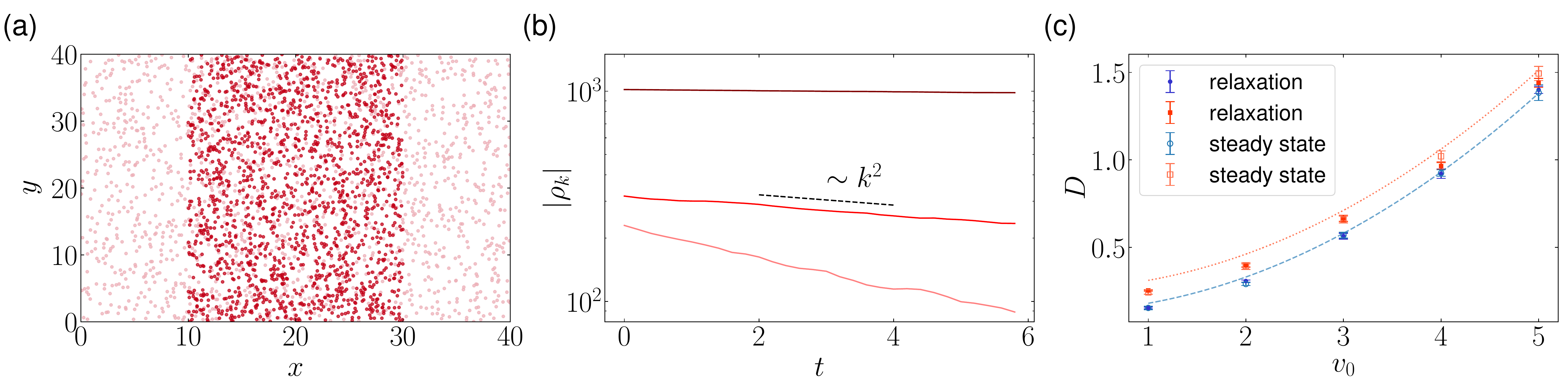}
	\caption{Density diffusion for interacting ABPs: (a) Macroscopic relaxation experiment where the particles are initialized to the middle half of the volume (red) and then allowed to evolve in time to a near homogenous configuration (pink). (b) The first three non-zero Fourier modes in the $x$-direction $|\rho_k|$ with $k \in \{ 2\pi/L,6\pi/L,10\pi/L \}$ (from top to bottom) display the expected exponential relaxation with a $k$-dependent slope confirming the hydrodynamic behavior. (c) Comparison of the transport coefficient measured using the macroscopic relaxation method (filled symbols) to the prediction of the Green-Kubo relation~\eqref{eq:GK_cons} obtained from steady-state correlation functions (open symbols) as a function of activity $v_0$ for two interaction strengths, $K=0.5$ (blue), and $K=1.0$ (red).  The dashed ($K = 0.5$) and dotted ($K = 1.0$) lines are analytic predictions from the linear theory. Parameters are in Appendix~\ref{sec:ABPs}.} 
	\label{fig:ABPs}
\end{figure*}

\subsection{Active brownian particles}
We first consider a fluid of $N$ spherical ABPs in a two-dimensional box of size $L\times L$, with periodic boundary conditions.
Interactions are modeled through a repulsive, short-ranged, pair potential $\phi(|{\bm r}_i-{\bm r}_j|)$.
Activity enters by each particle being self-propelled with a velocity $v_0\bm{e}(\theta_i) = v_0(\cos\theta_i, \sin\theta_i)$, whose orientation $\theta_i$ diffuses with diffusion coefficient $D_r$.
Including translational noise with diffusion coefficient $D_t$ leads to an evolution governed by the pair of overdamped Langevin equations~\cite{bialke2012crystallization,fily2012athermal}
\begin{equation}\label{eq:ABP}
\begin{aligned}
    \dot{\bm{r}}_i(t) 
    & = v_0 \bm{e}(\theta_i(t)) + \mu \bm{F}_i(t) + \sqrt{2D_t} \bm{\xi}_i(t), \\
    \dot{\theta}_i(t)
    & = \sqrt{2D_r} \eta_i(t),
\end{aligned}
\end{equation}
where ${\bm \xi}_i$ and $\eta_i$ are independent Gaussian white noises, $\mu$ is the bare mobility, and $\bm{F}_i = -\nabla_{\bm{r}_i} \sum_{j(\neq i)} \phi(|\bm{r}_i - \bm{r}_j|)$ is the the total force acting on the $i$-th particle due to pair interactions.

The only conserved variable is the total number of particles.
Momentum and energy are not conserved due to the self-propulsion and noise.
Accordingly, the only hydrodynamic variable is the local particle density $\rho_{\bm{r}}= \sum_{i} \delta(\bm{r} - \bm{r}_i)$, with steady-state average ${\bar \rho} = N/L^2$.
Since the particles do not prefer any particular direction, the density transport exhibits an unbiased isotropic diffusion with ${\bm v}=0$ and transport coefficient proportional to the identity matrix ${\mathsf D}=D{\mathsf I}$:
\begin{equation}\label{eq:macroscopic_ABPs}
    \partial_t\langle {\delta \rho}_{\bm{k}}(t) \rangle 
    = -k^2 D \langle \delta\rho_{\bm{k}}(t) \rangle,
\end{equation}
with a diverging relaxation time $\tau(k) = 1/(k^2D)$.

We first validate the Green-Kubo relation~\eqref{eq:GK_cons} through direct numerical simulation.
This is accomplished by first measuring the transport coefficient $D$ as the rate of exponential relaxation from an inhomogeneous initial condition per its definition and then comparing that to the prediction of the Green-Kubo relation.
As detailed in Appendix~\ref{sec:ABPs}, we use a harmonic interaction potential $\phi(r) = (K/2)(r - a)^2 \Theta(a - r)$, with interaction strength $K$ and length $a$, and we set $D_t=0$ to enhance the nonequilibrium effects and facilitate the numerical analysis.
To measure $D$ directly, we first observed the evolution of the Fourier modes $\rho_{\bm k}$ from a nonuniform initial condition with all the particles localized to the band $L/4 \leq x_i \leq 3L/4$, as depicted in Fig.~\ref{fig:ABPs}(a).
Figure \ref{fig:ABPs}(b) displays the expected exponential relaxation of a hydrodynamic mode whose slope is proportional to $k^2 D$.
Repeating this experiment 10 times and then averaging gives us our estimate of $D$ and its error (Fig.~\ref{fig:ABPs}(c)).
Next, using a single long steady-state simulation, we estimated $D$ from the steady-state correlation functions appearing in the Green-Kubo relation.
Figure \ref{fig:ABPs}(c) shows the coincidence of the two methods for measuring $D$ as a function of $v_0$ for two values of the interaction strength $K=\{0.5,1\}$, confirming the validatity of the Green-Kubo relation. 

To gain further insight, we complement the numerical validation with an analytic analysis.
The microscopic dynamics of the density $\rho_{\bm r}$ is given by a nonlinear stochastic differential equation with multiplicative noise, known as the Dean equation~\cite{dean1996langevin,demery2014generalized}.  
It is further coupled to the self-propulsion orientation through the polarization density ${\bm p}_{\bm r} = \sum_{i} {\bm e}(\theta_i)\delta({\bm r}-{\bm r}_i)$ as well as higher harmonics.
To make analytic progress we take the limit of a dense, weakly-interacting fluid.
Following~\cite{demery2014generalized}, we then linearize these equations about the steady state ($\bar\rho$ and $\bar{\bm p}=0$), and truncate the orientation harmonics at ${\bm p}_{\bm r}$, since higher harmonics do not contribute in the limit $k\to 0$.
The resulting linear Gaussian dynamics derived in Appendix~\ref{sec:ABPs} can be solved analytically allowing us to determine each term in the Green-Kubo relation, which are all diagonal (${\mathsf C} = C{\mathsf I}$, ${\mathsf E}=E{\mathsf I}$) with elements
\begin{equation}\label{eq:GK_lin_abp}
\begin{aligned}
\tilde\chi & = \frac{N[D_t+v_0^2/(2D_r)]}{D_t +\mu \bar\rho \phi_0+v_0^2/(2D_r)}, \\
C & = \frac{Nv_0^2}{2D_r}, ~~~  E = ND_t,
\end{aligned}
\end{equation}
where $\phi_0=\lim_{k\to 0}\phi_{\bm k}$ is the zero wave-vector limit of the pair potential.
Combining, we arrive at a prediction for the transport coefficient 
\begin{equation}\label{eq:D_lin_ABPs}
    D = D_t + \mu\bar{\rho}\phi_0 + \frac{v_0^2}{2D_r}.
\end{equation}
Weak-interactions and activity enhance the diffusion.
We further see for an equilibrium Brownian fluid where $v_0=0$, the current-current correlation function vanishes ($C=0$), and the transport coefficient is determined solely by $E$.
Thus in stochastic models even in equilibrium, ${\mathsf E}$ is required for an accurate prediction of the transport coefficient.

The predictions from this linearized theory~\eqref{eq:D_lin_ABPs} are compared to the simulation in Fig.~\ref{fig:ABPs}(c) where the dashed line is for the weaker interaction ($K=0.5$) and the dotted is the stronger interaction ($K=1.0)$.
For the weaker interaction ($K=0.5$), the linear approximation agrees well with the simulations when $v_0$ is large.
When $v_0$ is small, the predictions of the linear theory for both the strong and weak interactions fall outside the error bars, overestimating the transport coefficient.
Even still, the coincidence of the two numerical measurements of $D$ suggest the macroscopic Green-Kubo relation remains valid.
This illustrates how even when the microscopic dynamics are nonlinear, the emergent macroscopic dynamics can still be linear, as emphasized by Van Kampen~\cite{vankampen1969nonlinear}.

\begin{figure*}
	\includegraphics[width=2\columnwidth]{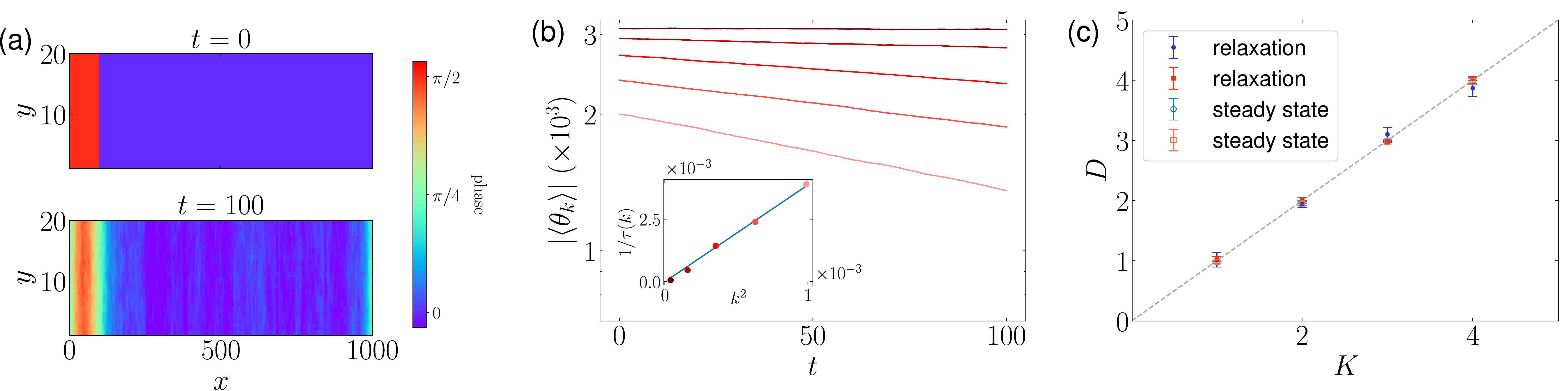}
	\caption{Phase diffusion in the synchronized Kuramoto model: (a) Macroscopic relaxation experiment where the left tenth of phases are initialized $\pi/2$ away from the homogenous state (top).  After $t=100$, the inhomogeneity has begun to diffuse away (bottom). (b) Plot of the first five smallest $k$  Fourier modes in the $x$-direction as a function of time, which verifies the expected exponential relaxation.  Their slope (inset) is linearly proportional to $k^2$ whose slope is a measurement of $D$. (c) Comparison of the transport coefficient measured using the macroscopic relaxation method (filled symbols) to the prediction of the Green-Kubo relation~\eqref{eq:GK_NG_modes} obtained from steady-state correlation functions (open symbols) as a function interaction strength $K$ for two noise strengths $(T, \sigma) =  (0.01, 0.01)$ (blue) and $(T, \sigma) = (0.002, 0.005)$ (red).}
\label{fig:kuramoto}
\end{figure*}

\subsection{Stochastic Kuramoto model}

Our second illustration is a variant of the Kuramoto model, which is a canonical model for synchronization among a large collection of phase oscillators~\cite{acebron2005kuramoto}.
In the original Kuramoto model, all oscillators interacted with each other and evolved deterministically~\cite{kuramoto1975self}, though later variations included a variety of modifications such as allowing for dynamical noise~\cite{sakaguchi1988cooperative}.
Here, we consider a noisy version with $N=L_xL_y$ oscillators evenly spaced on a two-dimensional square lattice of size $L_x\times L_y$ with nearest-neighbor interactions and periodic boundary conditions.
The time evolution of the $i$-th oscillator's phase $\theta_i$ is described by the stochastic equation
\begin{equation}\label{eq:Kuramoto_model}
    \dot{\theta}_i(t)
    = \Omega_i + K\sum_{j\sim i}\sin(\theta_j(t) - \theta_i(t))
    + \sqrt{2T} \xi_i(t),
\end{equation}
where the sum extends over neighbors of $i$, and $K$ is the interaction strength.
There are additionally two sources of noise. 
The oscillator's intrinsic frequency $\Omega_i$ is sampled from a Gaussian distribution of mean ${\bar \Omega}$ and variance $\sigma^2$.
Each oscillator also experiences white Gaussian dynamical noise $\xi_i$ with a strength characterized by $T$.
When the strength of noise, $\sigma$ and $T$, is weak and the interaction $K$ is strong, a finite system synchronizes and all the oscillators rotate coherently at a common frequency ${\bar \Omega}$.
The effect of the noise on this synchronization transition has recently been investigated in~\cite{sarkar2020noise}.

In the synchronized state, the rotational symmetry of the equations of motion~\eqref{eq:Kuramoto_model} is spontaneously broken.
As a result, we expect that localized perturbations in phase will relax diffusively via a Nambu-Goldstone mode.
We can see this explicitly if we consider a small perturbation to the synchronized state $\theta_i(t) = \theta_0 + \bar\Omega t+\delta\theta_i(t)$, where $\theta_0$ is an arbitrary offset that emerges when the symmetry is broken.
Linearizing the interaction in~\eqref{eq:Kuramoto_model} by assuming neighboring phases are close ($|\delta\theta_i - \delta\theta_j| \ll 1$), and Fourier transforming $(\theta_{\bm k}=\sum_j \theta_j e^{i {\bm k}\cdot{\bm r}_j}$), shows that each mode oscillates independently
\begin{equation}\label{eq:gaussian_kuramoto}
    \dot{\delta\theta}_{\bm{k}}(t)
    = \Omega_{\bm{k}} -k^2 K \delta\theta_{\bm{k}}(t) 
    + \sqrt{2T}\xi_{\bm{k}}(t),~~~ ({\bm k}\neq 0),
\end{equation}
for small $k$.
Upon averaging over the noises, we find that the average phase relaxes as
\begin{equation}\label{eq:linearized_Kuramoto}
    \partial_t\langle {\delta \theta}_{\bm{k}}(t) \rangle
    = -k^2 K \langle \delta\theta_{\bm{k}}(t) \rangle.
\end{equation}
In other words, the phase regresses diffusively with relaxation time $\tau(k)=1/(k^2 K)$ characterized by a transport coefficient $D=K$ that does not depend on the strength of the noises.

The transport coefficient $D$ can also be deduced via the Green-Kubo relation~\eqref{eq:GK_NG_modes}.
The exact calculation based on the linear equation~\eqref{eq:gaussian_kuramoto} gives for small $k$,
\begin{equation}
\begin{split}
\langle \delta\theta_{\bm{k}}\delta\theta_{-\bm{k}} \rangle_{\rm ss} &\simeq \frac{N\sigma^2}{k^4 K^2},\quad
 \langle \dot{\delta\theta}_{\bm{k}}\delta\theta_{-\bm{k}} \rangle_{\rm ss} \simeq -\frac{N\sigma^2}{k^2 K},\\
 & \langle \dot{\delta\theta}_{\bm{k}\omega}\dot{\delta\theta}_{-\bm{k}} \rangle_{\rm ss} \simeq\frac{N\sigma^2}{i\omega}.
 \end{split}
\end{equation}
Consequently, the Green-Kubo relation~\eqref{eq:GK_NG_modes} yields the consistent prediction $D = K$.
See Appendix~\ref{sec:Kuramoto} for more details on the linearized theory.

To corroborate this analysis, we estimated the transport coefficient by two independent numerical simulations.
We first setup an initial nonuniform profile of phase by rotating the leftmost tenth of oscillators ahead by $\pi/2$ and observed the subsequent regression of the first few smallest $k$ modes $\theta_{\bm{k}}$ in the $x$-direction, as illustrated in Fig.~\ref{fig:kuramoto}(a).
Figure~\ref{fig:kuramoto}(b) displays the expected linear regression of $\theta_{\bm{k}}$, averaged over 100 realizations, whose slope we use to extract $D=1/(k^2 \tau(k))$ (inset).
Steady-state simulations are then used to deduce $D$ from the correlation functions in the Green-Kubo relation~\eqref{eq:GK_NG_modes}.
Figure~\ref{fig:kuramoto}(c) shows the coincidence of the two different measurements of $D$ in comparison to the prediction from the linearized theory (dotted line).
As long as the strength of the noises is weak enough for the system to reach the synchronized state, the transport coefficient $D$ is only determined by $K$.
See Appendix~\ref{sec:Kuramoto} for thorough information about the simulation methods and the error analysis.

\section{Comparison with earlier results}\label{sec:earlier}

The near-equilibrium FDT and Green-Kubo relations are deep, long-standing results, and consequently have been extensively analyzed and extended in various directions.
In this section, we compare and contrast our present developments with pertinent earlier literature.
 
\subsection{Fluctuations and response}

In Sec.~\ref{sec:FRE}, we identified a class of dynamical perturbations whose response can be identified as a simple correlation function, akin to the equilibrium FDT.  
Graham has also identified a group of perturbations that result in an equilibrium-like fluctuation-response equality~\cite{graham1977covariant}.
Here, we elucidate the connection between these two approaches.

To make this connection, let us take a look at the steady-state distribution of our perturbed dynamics in Eq.~\eqref{eq:perturbed_equation} with external control parameter fixed $\lambda(t)={\bar \lambda}$.
The resulting steady-state distribution, given as the solution of ${\mathcal L}'P_{\rm ss}'({\bm z})=0$, is then approximately
\begin{equation}\label{eq:perturbed_ss}
P'_{\rm ss}({\bm z}) \simeq P_{\rm ss}({\bm z})\frac{1+\bar\lambda Q({\bm z})}{1+\bar\lambda \langle Q\rangle_{\rm ss}} \simeq \frac{P_{\rm ss}({\bf z}) e^{\bar\lambda Q({\bf z})}}{\left\langle e^{\bar\lambda Q}\right\rangle_{\rm ss}},
\end{equation}
accurate up to first order in ${\bar\lambda}$.
We see that the steady-state of the perturbed dynamics is approximately an exponential re-weighting of the unperturbed steady-state $P_{\rm ss}({\bm z})$ by the conjugate coordinate $Q$.

For nonequilibrium systems modeled using a Fokker-Planck equation,  Graham previously demonstrated that a perturbation that changes the steady-state via an exponential re-weighting as in Eq.~\eqref{eq:perturbed_ss} when applied dynamically will satisfy an equilibrium-like FDT, equivalent to Eq.~\eqref{eq:generalized_FDT}.
Graham's analysis, however, was performed by first deriving a covariant form of the Fokker-Planck equation and then identifying a decomposition of the dynamics where  it is more natural to work with exponential shifts in the steady-state distribution.
As noted in Ref.~\cite{eyink1996}, it is perhaps the rather formal mathematical language of Ref.~\cite{graham1977covariant} that has obscured this earlier prediction.

Our analysis complements this previous work in two important ways.  Our approach is valid for any Markovian dynamics and is not restricted to continuous-time continuous-space diffusion processes modeled via a Fokker-Planck equation
As such, we identify the pertinent perturbations using the original governing Master equation, without recourse to a covariant formulation.
This allows us to identify the changes in system parameters directly, as in Fig.~\ref{fig:ABP}.

\subsection{Green-Kubo expressions}

Green-Kubo relations for non-Hamiltonian dynamics -- such as hard-core classical fluids or systems described by stochastic Langevin equations -- have been deduced using the Mori-Zwanzig projection operator method independently by Ernst and Brito~\cite{ernst2005generalized,ernst2006new} as well as by Espa\~nol~\cite{espanol2002coarse,espanol2009einstein}.  While those authors explicitly only considered systems near equilibrium, their arguments trivially extend to homogenous nonequilibrium fluids like those considered here.

The work of Ernst and Brito most closely parallels the present analysis.  They projected directly onto the linear Langevin equation for the hydordynamic variables.
Espa\~nol, by contrast, employed the projection operator method to obtain the nonlinear Fokker-Planck equation describing the evolution of the full distribution of the hydrodynamic fluctuations.  
From this more general approach one can in principle obtain an equation for the linear regression of the hydrodynamic variables through a suitable expansion, as detailed for example by Zwanzig~\cite{zwanzig2001nonequilibrium,zwanzig1972}.
Thus, both analyses lead to similar conclusions.
In this section, we demonstrate how our linear response theory derivation of the Green-Kubo relations agrees with the relations obtained using the projection operator method.
For simplicity and clarity of presentation, we will assume in this section there is a single conserved hydrodynamic variable and $\langle {\bm j}_{\bm k}A_{-{\bm k}}\rangle=0$, so that ${\bm v}={\mathsf E}=0$.

To state the prediction obtained from the projection operator method, we first need to introduce a couple of concepts.
The first object we will need is the generator of the time-reversed~\cite{ernst2005generalized,ernst2006new,espanol2002coarse,espanol2009einstein} or dual dynamics~\cite{crooks1999,chernyak2006}
\begin{equation}
\tilde{\mathcal L}= P_{\rm ss}({\bm z}) {\mathcal L}^\dag P_{\rm ss}({\bm z})^{-1},
\end{equation}
where $\mathcal{L}^\dagger$ is the adjoint operator of $\mathcal{L}$ defined by $\int f(\bm{z})\mathcal{L}^\dagger g(\bm{z}) d\bm{z} = \int g(\bm{z})\mathcal{L} f(\bm{z}) d\bm{z}$.
We have also assumed that there are only even variables under time-reversal.
If there are odd variables, one must additionally reverse their sign by including the time-reversal operator (see Ref.~\cite{manzano2018} for an exposition in the quantum context).
In the dynamics generated by $\tilde{\mathcal L}$, every trajectory of the original dynamics appears with the same probability except run in reverse~\cite{risken2012fokker}.
When $\tilde{\mathcal L}={\mathcal L}$, the dynamics are said to be detailed balance, and every trajectory occurs with the same probability as its time-reverse~\cite{crooks1999}.
Apart from this interpretation, we will also find useful the following property of the dual generator,
\begin{equation}\label{eq:dualCommute}
{\mathcal L}(O({\bm z})P_{\rm ss}({\bm z}) ) = P_{\rm ss}({\bm z})\tilde{\mathcal L}^\dag O({\bm z}),
\end{equation}
 which allows us to ``commute'' the generator ${\mathcal L}$ past the steady-state distribution at the expense of introducing the adjoint of the dual generator.  
With the generator and its dual in hand, we can use them to define time-dependent hydrodynamic variables in the Heisenberg picture going forwards and backwards in time as
\begin{align}
&\partial_t \hat{A}_{\bm k}^H(t) ={\mathcal L}^\dag \hat{A}_{\bm k}^H(t)=i{\bm k}\cdot{\bm j}^H_{\bm k}(t)\\
&\partial_t \tilde{A}_{\bm k}^H(t) =-\tilde{\mathcal L}^\dag \tilde{A}_{\bm k}^H(t)=i{\bm k}\cdot \tilde{\bm j}^H_{\bm k}(t).
\end{align}
Here, the superscript $H$ stands for Heisenberg picture.  
We have also taken this opportunity to introduce local currents for both the forward ${\bm j}^H_{\bm k}$ and time-reversed $\tilde{\bm j}^H_{\bm k}$ dynamics.
This should be contrasted with the conservation equations introduced earlier, where the derivative was taken along a single dynamical trajectory.

The projection operator method then leads to the following Green-Kubo relation for the transport coefficient~\cite{ernst2005generalized,ernst2006new,espanol2002coarse,espanol2009einstein}
\begin{equation}\label{eq:projOp}
{\mathsf D}\tilde\chi = \lim_{k\to 0} \int_0^\infty dt \left\langle {\bm j}_{\bm k}^H(t) \tilde{\bm j}_{-{\bm k}}^H(0)\right\rangle_{P_{\rm ss}},
\end{equation}
where the subscript on the average $\langle O\rangle_{P_{\rm ss}}=\int O({\bm z})P_{\rm ss}({\bm z}) d{\bm z}$ emphasizes that this is a static average over the steady-state distribution as opposed to the average over steady-state dynamical trajectories employed in the rest of the paper, denoted by $\langle \cdot\rangle_{\rm ss}$. 
Much has been made over the fact that the correlation function in Eq.~\eqref{eq:projOp} contains a mixture of the forward and time-reversed dynamics~\cite{ernst2005generalized,ernst2006new,espanol2002coarse,espanol2009einstein}.
Indeed, it leads one to believe that to measure this correlation function one needs access to both the forward and the time-reversed dynamics in the same experimental setup.
Admittedly, this appears challenging in an experiment, though the effect of time-reversal can be readily be extracted computationally, as was done in Ref.~\cite{jung2016computing}.

The equality in Eq.~\eqref{eq:projOp} appears identical to our expression for the Green-Kubo relation \eqref{eq:GK_cons} except under the integral is a two-time average of Heisenberg operators going in two time directions as opposed to the forward-in-time current-correlation function in Eq.~\eqref{eq:GK_cons}.
In fact, these two expressions are mathematically identical, and Eq.~\eqref{eq:projOp} is simply the Heisenberg representation of a current correlation function.
To see this, observe that
\begin{equation}
\begin{aligned}
    & i{\bm k}\cdot \left\langle {\bm j}_{\bm k}^H(t) \tilde{\bm j}_{-{\bm k}}^H(0)\right\rangle_{P_{\rm ss}} \cdot i{\bm k} \\ 
    & =  \int d{\bm z} ~ \left[ e^{t{\mathcal L}^\dag}{\mathcal L}^\dag {\hat A}_{\bm k}({\bm z})\right]\left[\tilde {\mathcal L}^\dag \hat{A}_{-{\bm k}}({\bm z})\right] P_{\rm ss}({\bm z})\\
    & = \int d{\bm z} ~ {\hat A}_{\bm k}({\bm z})e^{t{\mathcal L}}{\mathcal L}^2\left[{\hat A}_{-{\bm k}}({\bm z})P_{\rm ss}({\bm z})\right],
\end{aligned}
\end{equation}
where in the last line we used Eq.~\eqref{eq:dualCommute}.
Next, we can recognize the resulting operator expression as the time derivative of the steady-state correlation function upon comparison with Eq.~\eqref{eq:corr}, to find
\begin{equation}
\begin{aligned}
    i{\bm k}\cdot \left\langle {\bm j}_{\bm k}^H(t) \tilde{\bm j}_{-{\bm k}}^H(0)\right\rangle_{P_{\rm ss}} \cdot i{\bm k}
    & = -\partial_t \partial_s \langle \hat{A}_{\bm k}(t) \hat{A}_{-{\bm k}}(s)\rangle_{\rm ss} \Big|_{s=0} \\
    & = i{\bm k}\cdot \left\langle {\bm j}_{\bm k}(t) {\bm j}_{-{\bm k}}(0)\right\rangle_{\rm ss}\cdot i{\bm k}. 
\end{aligned}
\end{equation}
Thus, the Green-Kubo relations obtained using the projection operator method contains the same current-correlation functions as obtained here, but expressed in terms of the Heisenberg picture.

When the dynamics are stochastic or not time-reversal symmetric, it is not well appreciated that when standard correlation functions are expressed in the Heisenberg picture they include the time-reversed or dual dynamics.  
This point may have interfered with the widespread understanding that the projection operator method when applied to the study of nonequilibrium fluids leads to predictions equivalent in form to equilibrium.

\section{Conclusion}

We have identified a class of perturbations whose response verifies an equilibrium-like fluctuation-response equality.
This allowed us to rationalize Onsager's regression hypothesis around nonequilibrium steady states and served as the foundation of a systematic method to extract linearized hydrodynamic transport equations around homogenous nonequilibrium steady-states.
The key resulting predictions were Green-Kubo relations that link hydrodynamic transport coefficients to steady-state fluctuations, which were verified in two models both analytically and numerically.
As for the traditional equilibrium Green-Kubo relations, our approach here complements results for non-Hamiltonian and stochastic systems based on the projection operator method~\cite{ernst2005generalized,ernst2006new,espanol2002coarse,espanol2009einstein}.
In the projection operator method, one must conjecture a criterion to single out a projected state that contains all relevant macroscopic dynamical information.
On the other hand, we make no assumptions about the microscopic distribution corresponding to the inhomogeneous state and thus avoid the choice of projected state; however, we do assume that our perturbation generates a response that captures the long-time correlations.
As a result, we naturally find Green-Kubo expressions in terms of simple steady-state correlation-functions.

Our derivation of Green-Kubo relations is valid for any system as long as the microscopic dynamics is Markovian and the steady-state is statistically translationally invariant.
Relaxing the Markovian assumption may be possible, since non-Markovian dynamics can be made Markovian by introducing auxiliary variables~\cite{gardiner2009stochastic}.
Another important direction is to inhomogenous boundary-driven steady-states, when the environmental interactions can be modeled via Markovian stochastic processes.

Finally, near-equilibrium time-reversal symmetry implies that cross-transport coefficients are equal, a prediction known as Onsager reciprocity~\cite{onsager1931reciprocal}.
The microscopic expressions for transport coefficients valid far-from-equilibrium derived here open the door to studying the violation of Onsager reciprocity as well as its connection to time-reversal-symmetry breaking and dissipation.

\section*{Acknowledgements}
We thank Suriyanarayanan Vaikuntanathan and Alexandre Solon for valuable discussions.
We also thank Freddy A. Cisneros for providing us the simulation code for the ABP simulations.

\appendix

\section{Equilibrium-like fluctuation-response relation}\label{sec:equivalence}
In this section, we derive and analyze the equilibrium-like fluctuation-response equality~\eqref{eq:generalized_FDT}. 

To proceed, we will solve to linear order the Master equation of the perturbed dynamics 
\begin{equation}\label{eq:perturbed_equation_appdx}
    \partial_t P(\bm{z},t)
    = \mathcal{L} P(\bm{z},t) -
    \lambda(t) \mathcal{L} ( Q(\bm{z}) P(\bm{z},t) ).
\end{equation}
We assume the unperturbed system ($\lambda = 0$) relaxes to a unique steady-state distribution $P_{\rm ss}(\bm{z})$ given as the solution of $\mathcal{L}P_{\rm ss}(\bm{z}) = 0$.
Taking the initial condition at $t=-\infty$ as the steady-state distribution $P(\bm{z},-\infty) = P_{\rm ss}(\bm{z})$, the formal solution of \eqref{eq:perturbed_equation_appdx} is
\begin{equation}
    P(\bm{z},t)
    = P_{\rm ss}(\bm{z}) - \int_{-\infty}^t ds ~ \lambda(s) e^{(t-s)
    \mathcal{L}} \mathcal{L}(Q({\bm z}) P_{\rm ss}(\bm{z})),
\end{equation}
up to linear order in the perturbation.
Accordingly, we arrive at Eq.~\eqref{eq:generalized_FDT} for the response of an observable
\begin{equation}\label{eq:response}
\begin{aligned}
    & \langle O(t) \rangle - \langle O \rangle_{\rm ss} \\
    & = -\int_{-\infty}^t ds ~ \lambda(s) \int d{\bm z}\ O(\bm{z}) e^{(t-s)\mathcal{L}} \mathcal{L}
    \left(Q(\bm{z}) P_{\rm ss}(\bm{z})\right) \\
    & = \int_{-\infty}^t ds ~ \lambda(s) \langle O(t) \dot{Q}(s) 
    \rangle_{\rm ss}.
\end{aligned}
\end{equation}
Here, we have identified the time-translationally-invariant steady-state correlation function, which is defined for any two observables, $O_1(\bm{z})$ and $O_2(\bm{z})$,  by
\begin{equation}
\begin{aligned}
    & \langle O_1(t) O_2(0) \rangle_{\rm ss} \\
    & = \int d{\bm z}d{\bm z}'\ O_1({\bm z})P({\bm z}, t|{\bm z}',0) O_2({\bm z}')P_{\rm ss}({\bm z}') \\
    & =\int d{\bm z}\ O_1({\bm z})e^{t{\mathcal L}} \left(O_2({\bm z})P_{\rm ss}({\bm z})\right)
\end{aligned}
\end{equation}
in terms of the transition probability $P({\bm z}, t|{\bm z}',0)$ obtained as the solution of \eqref{eq:perturbed_equation_appdx} with a delta-function initial condition $\delta({\bm z}-{\bm z}')$.

The defining characteristic of this perturbation is that the conjugate coordinate $Q({\bm z})$ appearing in the perturbation appears naturally in the response correlated with the observable.  Since this attribute is central to our analysis, we next deduce the class of perturbations with this property.

To this end, consider the dynamics in the presence of a generic linear perturbation
\begin{equation}\label{eq:perturbed_equation_for_M}
    \partial_t P(\bm{z},t)
    = \mathcal{L} P(\bm{z},t) -
    \lambda(t) \mathcal{M} P(\bm{z},t),
\end{equation}
for an arbitrary linear operator ${\mathcal M}$ suitably well behaved.
Following the identical linear perturbation theory analysis as above, we find that the response of an observable to this perturbation is
\begin{equation}\label{eq:FRR}
\begin{aligned}
    & \langle O(t) \rangle - \langle O \rangle_{\rm ss} \\
    & = -\int_{-\infty}^t ds ~ \lambda(s) \int d{\bm z}\ O(\bm{z}) e^{(t-s)\mathcal{L}} \mathcal{M}
    P_{\rm ss}(\bm{z}).
\end{aligned}
\end{equation}

By comparing Eqs.~\eqref{eq:response} and \eqref{eq:FRR}, we see that the condition that the linear response to the perturbation of ${\mathcal M}$ is identical to ${\mathcal L}(Q\cdot)$  is
\begin{equation}\label{eq:equivalence_cond1}
\begin{aligned}
    & \int_{-\infty}^t ds ~ \lambda(s) \int d\bm{z}~
    O(\bm{z}) e^{(t-s)\mathcal{L}} \mathcal{M} P_{\rm ss}(\bm{z}) \\
    & = \int_{-\infty}^t ds ~ \lambda(s) \int d\bm{z}~
    O(\bm{z}) e^{(t-s)\mathcal{L}} \mathcal{L} ( Q(\bm{z}) P_{\rm ss}(\bm{z}) ).
\end{aligned}
\end{equation}
To satisfy~\eqref{eq:equivalence_cond1} for an arbitrary observable $O$, the action of the generator $\mathcal{M}$ on the steady-state distribution must be the same as ${\mathcal L}(Q\cdot)$:
\begin{equation}\label{eq:equivalence_cond2}
    \mathcal{M} P_{\rm ss}(\bm{z}) = \mathcal{L} ( Q(\bm{z}) P_{\rm ss}(\bm{z})).
\end{equation}

In Sec.~\ref{sec:FRE}, we demonstrated that our dynamical perturbation \eqref{eq:perturbed_equation} is equivalent to perturbing the energy of a deterministic Hamiltonian dynamics.
We can now use our equivalence condition \eqref{eq:equivalence_cond2} to demonstrate that perturbing the potential energy of an equilibrium system generates the same response as our perturbation~\eqref{eq:perturbed_equation} for a broader class of nondeterministic dynamics.

To this end, we consider a general underdamped Langevin dynamics whose generator is given by (arguments are suppressed for clarity)
\begin{equation}\label{eq:underdamped_Langevin}
\begin{aligned}
    \mathcal{L}P
    & = - (\bm{\nabla}_{\bm{p}} H) \cdot (\bm{\nabla}_{\bm{q}} P)
    + (\bm{\nabla}_{\bm{q}} H) \cdot (\bm{\nabla}_{\bm{p}} P) \\
    & ~~~ - \bm{\nabla}_{\bm{p}} \cdot \left( \bm{f}_{\rm nc} P - \mathsf{G} \cdot \mathbf{p} P - \mathsf{T} \cdot (\bm{\nabla}_{\bm{p}} P) \right)
\end{aligned}
\end{equation}
where $\bm{q}$ and $\bm{p}$ are collective notations for the positions and momenta of all constituent particles, $H = H(\bm{q},\bm{p})$ is a Hamiltonian-like energy function, $\bm{f}_{\rm nc}$ is a nonconservative force, $\mathsf{G}$ is a matrix describing friction, and $\mathsf{T}$ is a matrix describing thermal fluctuations.
The underdamped Langevin dynamics~\eqref{eq:underdamped_Langevin} encompasses (i) Hamiltonian dynamics as a special case where $\bm{f}_{\rm nc} = \mathsf{G} = \mathsf{T} = 0$ and (ii) overdamped Langevin dynamics as a limiting case where friction and thermal fluctuation are strong.

We define a potential energy perturbation as a change of the energy function $H(\bm{z}) \to H(\bm{z}) - \lambda(t) U(\bm{q})$, whose generator is given by $\mathcal{L} - \lambda(t) \mathcal{M}$ with
\begin{equation}
    \mathcal{M} P(\bm{z},t) = (\bm{\nabla}_{\bm{q}} U(\bm{q}) ) \cdot (\bm{\nabla}_{\bm{p}} P(\bm{z},t)).
\end{equation}
Let us compare this with our perturbation \eqref{eq:perturbed_equation} with the choice $Q({\bf z})=U({\bf q})$:
\begin{equation}
    \mathcal{L} ( U(\bm{q}) P_{\rm ss}(\bm{z},t) )
    = -P_{\rm ss}(\bm{z}) (\bm{\nabla}_{\bm{p}} H(\bm{z}) ) \cdot (\bm{\nabla}_{\bm{q}} U(\bm{q})),
\end{equation}
The equivalence condition~\eqref{eq:equivalence_cond2} then implies the potential energy perturbation is equivalent to $\mathcal{L} - \lambda(t) \mathcal{L} U(\bm{q})$ if
\begin{equation}\label{eq:equivalence_cond3}
    \bm{\nabla}_{\bm{p}} P_{\rm ss}(\bm{z}) = -P_{\rm ss}(\bm{z})\bm{\nabla}_{\bm{p}} H(\bm{z}),
\end{equation}
or equivalently, $\ln P_{\rm ss}(\bm{z}) = -H(\bm{z}) + h(\bm{q})$ for any arbitrary position-dependent function $h(\bm{q})$.
The condition~\eqref{eq:equivalence_cond3} is satisfied at equilibrium where the steady-state distribution is given by the Boltzmann distribution $P_{\rm ss}({\bf z}) = e^{- H(\bm{z})}/(\int d\bm{z} e^{- H(\bm{z})})$ in units of $\beta = 1$.
Therefore, the potential energy perturbation $H(\bm{z}) \to H(\bm{z}) - \lambda(t) U(\bm{q})$ is equivalent to $\mathcal{L} - \lambda(t) \mathcal{L} U(q)$ at equilibrium.

\section{Derivation of Green-Kubo relations}\label{sec:gen_theory}
In this section, we elaborate the connection between the equilibrium-like fluctuation-response equality and Green-Kubo relations in the presence of multiple hydrodynamic variables.

Consider a system with $n$ hydrodynamic variables, $\{ \hat{A}_{\bm{r}}^{\alpha}(\bm{z}) \}_{\bm{r}\in V}$ with $\alpha = 1,\cdots,n$, locally defined at each location $\bm{r}$.
In a homogeneous steady state, the hydrodynamic variables have uniform mean values $\bar{A}^\alpha = \langle A_{\bm{r}}^\alpha \rangle_{\rm ss}$.
Hydrodynamic transport coefficients describe the rate at which an inhomogeneity relaxes away as the system approaches the homogeneous steady state.
In order to exploit~\eqref{eq:generalized_FDT} to derive Green-Kubo relations for hydrodynamic transport coefficients, we choose the form of the perturbation functions to be
\begin{equation}
    Q(\bm{z}) = \int_V d\bm{r} ~ \hat{A}_{\bm{r}}^\alpha(\bm{z}) f_{\bm{r}}^\alpha, ~~~
    \lambda(t) = e^{\epsilon t}\Theta(-t),
\end{equation}
where $\epsilon$ is a positive constant which we will eventually take to zero so that that perturbation is turned on slowly and $\Theta(t)$ is the Heaviside step function forcing the perturbation off at $t=0$.
The conjugate fields $f^\alpha_{\bm{r}}$ are arbitrary as long as all integrals converge.
The Einstein summation convention is adapted for repeated Greek indices ($1\leq \alpha,\beta,\gamma \leq n$) throughout this section.
Under these choices, equality \eqref{eq:generalized_FDT} for $O = \delta\hat{A}^\alpha_{\bm{r}} = \hat{A}_{\bm{r}}^\alpha - \langle \hat{A}_{\bm{r}}^\alpha \rangle_{\rm ss}$ implies
\begin{equation}
\begin{aligned}
    & \langle \delta \hat{A}_{\bm{r}}^\alpha(t) \rangle
    = \langle \hat{A}_{\bm{r}}^\alpha(t) \rangle - \langle \hat{A}_{\bm{r}}^\alpha \rangle_{\rm ss} \\
    & = \int_{-\infty}^0 ds \int_V d\bm{r}' ~ e^{\epsilon s} \partial_s \langle \hat{A}_{\bm{r}}^\alpha(t) \hat{A}_{\bm{r}'}^\beta(s) 
    \rangle_{\rm ss} f_{\bm{r}'}^\beta
\end{aligned}
\end{equation}
for $t \geq 0$.
We eliminate the time integral by first integrating by parts and then taking the limit $\epsilon \to 0$ (adiabatically turned-on perturbation),
\begin{equation}
    \langle \delta \hat{A}_{\bm{r}}^\alpha(t) \rangle
    = \int_V d\bm{r}' ~ \langle \hat{A}_{\bm{r}}^\alpha(t) \hat{A}_{\bm{r}'}^\beta(s) 
    \rangle_{\rm ss} f_{\bm{r}'}^\beta.
\end{equation}
Using the translation invariance of a homogeneous steady state $\langle \hat{A}_{\bm{r}}^\alpha(t)  \hat{A}_{\bm{r}'}^\beta(0) \rangle_{\rm ss} = \langle \hat{A}_{\bm{r}-\bm{r}'}^\alpha(t) \hat{A}_{\bm{0}}^\beta(0) \rangle_{\rm ss}$ and an exchange of integration range $\int_V d\bm{r} \int_V d\bm{r}' \approx \int_V d(\bm{r}-\bm{r}') \int_V d\bm{r'}$ for a large system size $V$ twice, we obtain a compact symmetric expression for the Fourier modes $\hat{A}_{\bm{k}}^\alpha = \int_V d\bm{r} \hat{A}_{\bm{r}}^\alpha e^{i\bm{k}\cdot\bm{r}}$ as
\begin{equation}\label{eq:simple_FDT}
\begin{aligned}
    & \langle \delta \hat{A}_{\bm{k}}^\alpha(t) \rangle \\
    & \approx \int_V d(\bm{r}-\bm{r}') ~ 
    \langle \hat{A}_{\bm{r}-\bm{r}'}^\alpha(t)
    \hat{A}_{\bm{0}}^\beta(0) \rangle_{\rm ss} e^{i\bm{k}\cdot(\bm{r}-\bm{r}')} \\
    & ~~~ \times \int_V d\bm{r}' ~ f_{\bm{r}'}^\beta e^{i\bm{k}\cdot \bm{r}'}  \\
    & \approx
    \frac{1}{V} \int_V d\bm{r} \int_V d\bm{r}' 
    \langle \hat{A}_{\bm{r}}^\alpha(t)
    \hat{A}_{\bm{r}'}^\beta(0) \rangle_{\rm ss} e^{i\bm{k}\cdot(\bm{r}-\bm{r}')} 
    f_{\bm{k}}^\beta\\
    & = \frac{1}{V} \langle \hat{A}_{\bm{k}}^\alpha(t) \hat{A}_{-\bm{k}}^\beta(0) \rangle_{\rm ss} f_{\bm{k}}^\beta.
\end{aligned}
\end{equation}
The choice of $f_{\bm{k}}^\beta$ is immaterial since it can be eliminated by the initial condition $\langle \delta \hat{A}_{\bm{k}}^\alpha(0) \rangle = \langle \hat{A}_{\bm{k}}^\alpha(0) \hat{A}_{-\bm{k}}^\beta(0) \rangle_{\rm ss} f_{\bm{k}}^\beta / V$.

Although equality~\eqref{eq:simple_FDT} is based on the particular ensemble generated at $t=0$, the long-time behavior of the hydrodynamic modes is insensitive to this initial ensemble.
At the same time, the long length- and time-scale relaxation behavior of hydrodynamic variables $\delta A_{\bm{k}}^\alpha = \langle \delta \hat{A}_{\bm{k}}^\alpha \rangle$ towards the homogeneous steady state will be well captured by the linear regression equation
\begin{equation}\label{eq:linear_regression}
    \partial_t \delta A_{\bm{k}}^\alpha(t)
    = - M_{\bm{k}}^{\alpha\beta} \delta A_{\bm{k}}^\beta(t)
\end{equation}
with a generalized transport coefficient $M_{\bm{k}}$.
The connection between microscopic and macroscopic dynamics can be made by noticing that particularities of the initial ensemble at $t=0$ will die out fast and \eqref{eq:simple_FDT} should become universal at long length- and time-scales.
Therefore, comparing \eqref{eq:simple_FDT} and \eqref{eq:linear_regression}, we conclude that the equality
\begin{equation}\label{eq:Onsager_regression}
    \partial_t \langle \hat{A}_{\bm{k}}^\alpha(t) \hat{A}_{-\bm{k}}^\gamma(0) \rangle_{\rm ss}
    = -M_{\bm{k}}^{\alpha\beta} \langle \hat{A}_{\bm{k}}^\beta(t) \hat{A}_{-\bm{k}}^\gamma(0) \rangle_{\rm ss}
\end{equation}
holds at long times for small $k=|\bm{k}|$.

In the following, we make use of the procedure to drive Green-Kubo relations for equilibrium transport established by Oppenheim and collaborators~\cite{selwyn1971generalized,weare1974nonlinear}.
We begin by manipulating \eqref{eq:Onsager_regression} into a form that makes as many of the time-derivates explicit as possible, allowing us to do an expansion for long times and small $k$.
To this end, we note that the long-time-scale dynamical information is contained in the small frequency modes of the half-Fourier transform $\delta \hat{A}_{\bm{k}\omega} = \int_0^\infty dt e^{i\omega t} \hat{A}_{\bm{k}}(t)$.
Taking the half-Fourier transform in time, we can express~\eqref{eq:Onsager_regression} in two ways.
First, by directly taking the half-Fourier transform 
\begin{equation}\label{eq:transformed_FDT1}
    \langle \dot{\hat{A}}_{\bm{k}\omega}^\alpha \hat{A}_{-\bm{k}}^\gamma(0) \rangle_{\rm ss} 
    = -M_{\bm{k}}^{\alpha\beta} \langle \hat{A}_{\bm{k}\omega}^\beta \hat{A}_{-\bm{k}}^\gamma(0) \rangle_{\rm ss},
\end{equation}
or, second, by using the transform to replace a time derivative with a frequency as
\begin{equation}\label{eq:transformed_FDT2}
\begin{aligned}
    & -i\omega \langle \hat{A}_{\bm{k}\omega}^\alpha \hat{A}_{-\bm{k}}^\gamma(0) \rangle_{\rm ss}
    + M_{\bm{k}}^{\alpha\beta} 
    \langle \hat{A}_{\bm{k}\omega}^\beta \hat{A}_{-\bm{k}}^\gamma(0) \rangle_{\rm ss} \\
    & = \langle \hat{A}_{\bm{k}}^\alpha \hat{A}_{-\bm{k}}^\gamma \rangle_{\rm ss}.
\end{aligned}
\end{equation}
For brevity, from now on, we will suppress the time argument $t=0$ unless especially necessary for clarity, thus $\hat{A}_{-\bm{k}}^\gamma(0) = \hat{A}_{-\bm{k}}^\gamma$ so on.
Next, we multiply~\eqref{eq:transformed_FDT2} by $M_{\bm k}^{\alpha\beta}$ and then substitute \eqref{eq:transformed_FDT1}, to obtain
\begin{equation}\label{eq:transformed_FDT3}
\begin{aligned}
    & -i\omega \langle \dot{\hat{A}}_{\bm{k}\omega}^\alpha \hat{A}_{-\bm{k}}^\gamma \rangle_{\rm ss}  + M_{\bm{k}}^{\alpha\beta}
    \langle \dot{\hat{A}}_{\bm{k}\omega}^\beta \hat{A}_{-\bm{k}}^\gamma \rangle_{\rm ss} \\
    & = -M_{\bm{k}}^{\alpha\beta} \langle \hat{A}_{\bm{k}}^\beta \hat{A}_{-\bm{k}}^\gamma \rangle_{\rm ss}.
\end{aligned}
\end{equation}
Then we remove the frequency via the identities (i) $-i\omega \langle \dot{\hat{A}}_{\bm{k}\omega}^\alpha \hat{A}_{-\bm{k}}^\gamma \rangle_{\rm ss} = \langle \dot{\hat{A}}_{\bm{k}}^\alpha \hat{A}_{-\bm{k}}^\gamma \rangle_{\rm ss} - \langle \dot{\hat{A}}_{\bm{k}\omega}^\alpha \dot{\hat{A}}_{-\bm{k}}^\gamma \rangle_{\rm ss}$, which can be proved by integration by parts, and further utilizing the stationarity (ii) $\langle \dot{\hat{A}}_{\bm{k}\omega}^\beta \hat{A}_{-\bm{k}}^\gamma \rangle_{\rm ss} = - \langle \hat{A}_{\bm{k}\omega}^\beta \dot{\hat{A}}_{-\bm{k}}^\gamma \rangle_{\rm ss}$. 
Equation~\eqref{eq:transformed_FDT3} then becomes
\begin{equation}\label{eq:transformed_FDT4}
\begin{aligned}
    & \langle \dot{\hat{A}}_{\bm{k}}^\alpha \hat{A}_{-\bm{k}}^\gamma \rangle_{\rm ss} - \langle (\dot{\hat{A}}_{\bm{k}\omega}^\alpha + M_{\bm{k}}^{\alpha\beta} \hat{A}_{\bm{k}\omega}^\beta) \dot{\hat{A}}_{-\bm{k}}^\gamma \rangle_{\rm ss} \\
    & = -M_{\bm{k}}^{\alpha\beta} \langle \hat{A}_{\bm{k}}^\beta \hat{A}_{-\bm{k}}^\gamma \rangle_{\rm ss}.
\end{aligned}
\end{equation}
from which we can derive Green-Kubo relations for hydrodynamic transport coefficients.

It is useful to define a dissipative rate $\dot{\hat{B}}_{\bm{k}}^\alpha(t) = \dot{\hat{A}}_{\bm{k}}^\alpha(t) + M_{\bm{k}}^{\alpha\beta} \hat{A}_{\bm{k}}^\beta(t)$.
From equality~\eqref{eq:transformed_FDT1}, the correlation $\langle \dot{\hat{B}}_{\bm{k}\omega}^\alpha \hat{A}_{-\bm{k}}^\gamma \rangle_{\rm ss}$ vanishes in the hydrodynamic limit, suggesting that the fluctuations of $\dot{\hat{B}}_{\bm{k}}^\alpha$ decay on microscopic times.
Thus, the dissipative rate $\dot{\hat{B}}_{\bm{k}}^\alpha$ represents the rapidly fluctuating part of $\dot{\hat{A}}_{\bm{k}}$.
Since $\langle \dot{\hat{B}}_{\bm{k}\omega}^\alpha \dot{\hat{A}}_{-\bm{k}}^\gamma \rangle_{\rm ss} = \langle \dot{\hat{B}}_{\bm{k}\omega}^\alpha (\dot{\hat{B}}_{-\bm{k}}^\gamma - M_{-\bm{k}}^{\gamma\beta} \hat{A}_{-\bm{k}}^\beta) \rangle_{\rm ss} = \langle \dot{\hat{B}}_{\bm{k}\omega}^\alpha \dot{\hat{B}}_{-\bm{k}}^\gamma \rangle_{\rm ss}$, we can alternatively express~\eqref{eq:transformed_FDT4} in the compact symmetric form
\begin{equation}\label{eq:Onsager_hypothesis_final}
    \langle \dot{\hat{A}}_{\bm{k}}^\alpha \hat{A}_{-\bm{k}}^\gamma \rangle_{\rm ss}
    - \langle \dot{\hat{B}}_{\bm{k}\omega}^\alpha \dot{\hat{B}}_{-\bm{k}}^\gamma \rangle_{\rm ss}
    = -M_{\bm{k}}^{\alpha\beta} \langle \hat{A}_{\bm{k}}^\beta \hat{A}_{-\bm{k}}^\gamma \rangle_{\rm ss}.
\end{equation}
More generally, time-retardation effects in the macroscopic dynamics by allowing the generalized transport coefficient to frequency dependent $M_{\bm{k}\omega}$~\cite{forster1975hydrodynamic}.
Repeating the same algebraic procedure in this case leads to the same end result~\eqref{eq:Onsager_hypothesis_final} except for the replacement of $M_{\bm{k}}^{\alpha\beta}$ by $M_{\bm{k}\omega}^{\alpha\beta}$.
For simplicity, we adhere to the frequency-independent case.

The last step in the derivation of Green-Kubo relations is to connect hydrodynamic transport coefficients to the microscopic correlation functions using equality~\eqref{eq:Onsager_hypothesis_final}.
To this end, we assume that the generalized transport coefficient can be expanded as a series in $i\bm{k}$ in the hydrodynamic limit, which is equivalent to a gradient expansion in real space.
The first two (multi-component) coefficients $\bm{v}$ and $\mathsf{D}$ of the expansion $M_{\bm{k}}^{\alpha\beta} \approx -i\bm{k} \cdot \bm{v}^{\alpha\beta} + \bm{k} \cdot \mathsf{D}^{\alpha\beta} \cdot \bm{k} + \cdots$ describe the drift and diffusive behavior of the macroscopic dynamics, respectively.
Inserting this expansion of $M_{\bm{k}}$ into~\eqref{eq:Onsager_hypothesis_final} and comparing the terms order by order in $\bm{k}$, lead to the Green-Kubo relations.
Although this procedure can be done generally without any assumption on the hydrodynamic variables $\hat{A}_{\bm{k}}^\alpha$, it can be simplified for local densities of conserved variables thanks to corresponding continuity equations.
Thus, we derive Green-Kubo relations for local densities and Nambu-Goldstone modes separately.

\subsection{Local densities}
Local densities of conserved variables satisfy continuity equations $\partial_t \hat{A}_{\bm{r}}^\alpha = -\nabla \cdot \bm{j}_{\bm{r}}^\alpha$, which define the corresponding local currents $\bm{j}_{\bm{r}}^\alpha$.
The continuity equations in Fourier space, $\dot{\hat{A}}_{\bm{k}}^\alpha = i\bm{k} \cdot \bm{j}_{\bm{k}}^\alpha$, make explicit that the rates of change of local densities are at least linear in $\bm{k}$.
The dissipative rates $\dot{\hat{B}}_{\bm{k}}^\alpha = \dot{\hat{A}}_{\bm{r}}^\alpha + M_{\bm{k}}^{\alpha\beta} \hat{A}_{\bm{k}}^\beta \approx i\bm{k} \cdot (\bm{j}_{\bm{k}}^\alpha - \bm{v}^{\alpha\beta} \hat{A}_{\bm{k}}^\beta) + \cdots $ are also proportional to $\bm{k}$ in the small-$k$ limit, which suggests the definition of dissipative currents $\bm{I}_{\bm{k}}^\alpha = \bm{j}_{\bm{k}}^\alpha - \bm{v}^{\alpha\beta} \hat{A}_{\bm{k}}^\beta$~\cite{selwyn1971generalized,weare1974nonlinear}.
Plugging the definitions of currents into~\eqref{eq:Onsager_hypothesis_final}, leads to
\begin{equation}\label{eq:Onsager_hypothesis_conserved}
\begin{aligned}
    & i\bm{k} \cdot \langle \bm{j}_{\bm{k}}^\alpha \hat{A}_{-\bm{k}}^\gamma \rangle_{\rm ss}
    - \bm{k} \cdot \langle \bm{I}_{\bm{k}\omega}^\alpha \bm{I}_{-\bm{k}}^\gamma \rangle_{\rm ss} \cdot \bm{k} \\
    & = (i\bm{k} \cdot \bm{v}^{\alpha\beta} - \bm{k} \cdot \mathsf{D}^{\alpha\beta} \cdot \bm{k} + \cdots ) \langle \hat{A}_{\bm{k}}^\beta \hat{A}_{-\bm{k}}^\gamma \rangle_{\rm ss}.
\end{aligned}
\end{equation}
Defining the static correlation function as $\tilde{\chi}^{\beta\gamma} = \lim_{k\to0} \langle \hat{A}_{\bm{k}}^\beta \hat{A}_{-\bm{k}}^\gamma \rangle_{\rm ss}$ and comparing linear terms in $\bm{k}$, we find for $\bm{v}^{\alpha\beta}$
\begin{equation}
    \bm{v}^{\alpha\beta}\tilde{\chi}^{\beta\gamma} 
    = \lim_{k\to 0} \langle \bm{j}_{\bm{k}}^\alpha \hat{A}_{-\bm{k}}^\gamma \rangle_{\rm ss}.
\end{equation}
The dissipative-current correlation term $\bm{k} \cdot \langle \bm{I}_{\bm{k}\omega}^\alpha \bm{I}_{-\bm{k}}^\gamma \rangle_{\rm ss} \cdot \bm{k}$ does not appear, since it is at least quadratic in $\bm{k}$.
By introducing a first-order correction matrix $\mathsf{E}^{\alpha\gamma}$ via
\begin{equation}\label{eq:GK_v_density}
    \langle \bm{j}_{\bm{k}}^\alpha \hat{A}_{-\bm{k}}^\gamma \rangle_{\rm ss}
    = \bm{v}^{\alpha\beta}\tilde{\chi}^{\beta\gamma} + \mathsf{E}^{\alpha\gamma} \cdot i\bm{k} + \cdots,
\end{equation}
we can also derive a Green-Kubo relation for $\mathsf{D}^{\alpha\beta}$ by comparing second order terms,
\begin{equation}\label{eq:GK_D_density}
    \mathsf{D}^{\alpha\beta} \tilde{\chi}^{\beta\gamma}
    = \lim_{\omega\to 0}\lim_{k \to 0} \langle \bm{I}_{\bm{k}\omega}^\alpha \bm{I}_{-\bm{k}}^\gamma \rangle_{\rm ss} + \mathsf{E}^{\alpha\gamma}.
\end{equation}

\subsection{Nambu-Goldstone modes}
Nambu-Goldstone modes, which emerge when a continuous symmetry is broken, are not necessarily conserved.
Thus the corresponding local currents are not defined in general due to the absence of continuity equations, and thus the $k$-dependence of the rate $\dot{\hat{A}}_{\bm{k}}^\alpha$ is not simple. 
Moreover, steady states with a broken continuous symmetry are characterized by a long-range correlation, which is indicated by the divergence of the limit $\lim_{k\to0} \langle \hat{A}_{\bm{k}}^\beta \hat{A}_{-\bm{k}}^\gamma \rangle_{\rm ss}$.
Due to the absence of local currents and finite static correlation functions, which made the derivations for local densities simpler, we need to consider the small-$k$ expansion of~\eqref{eq:Onsager_hypothesis_final} more generally.
Thus, we take a divergence of the form $\langle \hat{A}_{\bm{k}}^\beta \hat{A}_{-\bm{k}}^\gamma \rangle_{\rm ss} \sim k^{-q}$ with an arbitrary exponent $q$.
We recall~\eqref{eq:transformed_FDT4} here for convenience:
\begin{equation}
\begin{aligned}
    & \langle \dot{\hat{A}}_{\bm{k}}^\alpha \hat{A}_{-\bm{k}}^\gamma \rangle_{\rm ss} - \langle (\dot{\hat{A}}_{\bm{k}\omega}^\alpha + M_{k}^{\alpha\beta} \hat{A}_{\bm{k}\omega}^\beta) \dot{\hat{A}}_{-\bm{k}}^\gamma \rangle_{\rm ss} \\
    & = -M_{k}^{\alpha\beta} \langle \hat{A}_{\bm{k}}^\beta \hat{A}_{-\bm{k}}^\gamma \rangle_{\rm ss}
\end{aligned}
\end{equation}
To be consistent with macroscopic regression, the leading orders must balance in the small-$k$ limit.
Therefore, comparing the leading order terms of $k$ after taking the following considerations, we obtain a Green-Kubo relation for ${\bm v}^{\alpha\beta}$:
First, the term $M_{\bm{k}}^{\alpha\beta} \langle \hat{A}_{\bm{k}\omega}^\beta \dot{\hat{A}}_{-\bm{k}}^\gamma \rangle_{\rm ss} = M_{\bm{k}}^{\alpha\beta} M_{\bm{k}}^{\beta\delta} \langle \hat{A}_{\bm{k}\omega}^\delta \hat{A}_{-\bm{k}}^\gamma \rangle_{\rm ss}$ is clearly a higher-order correction.
Second, in contrast to local densities for conserved quantities, there is no guarantee that $\langle \dot{\hat{A}}_{\bm{k}\omega}^\alpha \dot{\hat{A}}_{-\bm{k}}^\gamma \rangle_{\rm ss}$ is higher-order than $\langle \dot{\hat{A}}_{\bm{k}\omega}^\alpha \hat{A}_{-\bm{k}}^\gamma \rangle_{\rm ss}$.
Lastly, we define inverse matrix $[\langle \hat{A}_{\bm{k}} \hat{A}_{-\bm{k}} \rangle_{\rm ss}]^{-1}$ such that $\langle \hat{A}_{\bm{k}}^\beta \hat{A}_{-\bm{k}}^\gamma \rangle_{\rm ss} \cdot [\langle \hat{A}_{\bm{k}} \hat{A}_{-\bm{k}} \rangle_{\rm ss}]^{-1}_{\gamma\alpha} = \delta_{\alpha\beta}$.
Consequently, we have
\begin{widetext}
\begin{equation}\label{eq:GK_v_NG}
    {\hat{\bm k}}\cdot{\bm v}^{\alpha\beta} = \lim_{\omega\to 0}\lim_{k\to 0}
    \frac{1}{ik} \left( \langle \dot{\hat{A}}_{\bm{k}}^\alpha \hat{A}_{-\bm{k}}^\gamma \rangle_{\rm ss} - \langle \dot{\hat{A}}_{\bm{k}\omega}^\alpha \dot{\hat{A}}_{-\bm{k}}^\gamma \rangle_{\rm ss}\right) \cdot \left[ \langle \hat{A}_{\bm{k}} \hat{A}_{-\bm{k}} \rangle_{\rm ss} \right]^{-1}_{\gamma \beta}
\end{equation}
with $\hat{\bm{k}} = \bm{k}/k$.
The next order correction leads to a Green-Kubo relation for $D^{\alpha\beta}$,
\begin{equation}\label{eq:GK_D_NG}
    {\hat{\bm k}}\cdot{\mathsf D}^{\alpha\beta}\cdot{\hat{\bm k}}
    = \lim_{\omega\to 0}\lim_{k\to 0}
    \frac{1}{k^2} \left( \langle (\dot{\hat{A}}_{\bm{k}\omega}^\alpha - ikv^{\alpha\beta} \hat{A}_{\bm{k}\omega}^\beta) \dot{\hat{A}}_{-\bm{k}}^\gamma \rangle_{\rm ss}
    - \langle (\dot{\hat{A}}_{\bm{k}}^\alpha -ik v^{\alpha\beta} \hat{A}_{\bm{k}}^\beta ) \hat{A}_{-\bm{k}}^\gamma \rangle_{\rm ss}
    \right) \cdot \left[ \langle \hat{A}_{\bm{k}} \hat{A}_{-\bm{k}} \rangle_{\rm ss} \right]^{-1}_{\gamma \beta}
\end{equation}
\end{widetext}
It is noticeable that \eqref{eq:GK_v_NG} and \eqref{eq:GK_D_NG} are reduced to \eqref{eq:GK_v_density} and \eqref{eq:GK_D_density} when the mode is conserved.
In fact, \eqref{eq:GK_v_NG} and \eqref{eq:GK_D_NG} are generally valid for any hydrodynamic variables, since they are derived without any prior knowledge about $\hat{A}_{\bm{k}}^\alpha$.

\section{Active Brownian Particles}\label{sec:ABPs}
\subsection{Linearized Dean equation}
In this section, we derive a coupled linearized Dean equation for a system of active Brownian particles.
For the sake of readability, time arguments are suppressed in this section unless strictly necessary.
The resulting linear equations predict a Green-Kubo relation for the diffusive transport coefficient $D$ for the case of soft interactions at a high density.
We recall the equations of motion for active Brownian particles~\eqref{eq:ABP}:
\begin{equation}
\begin{aligned}
    \dot{\bm{r}}_i(t)
    & = v_0 \bm{e}(\theta_i(t)) + \mu \bm{F}_i(t) + \sqrt{2D_t} \bm{\xi}_i(t), \\
    \dot{\theta}_i(t)
    & = \sqrt{2D_r} \eta_i(t),
\end{aligned}
\end{equation}
where $\bm{F}_i = -\nabla_{\bm{r}_i} \sum_{j\neq i} \phi(|\bm{r}_i - \bm{r}_j|)$ with interaction potential $\phi(r)$; simulations are performed with the specific choice $\phi(r) = (K/2)(r - a)^2 \Theta(a - r)$.
The time evolution equation of the local particle number density $\rho_{\bm{r}} = \sum_{j=1}^N \delta(\bm{r}-\bm{r}_j)$, known as the Dean equation~\cite{dean1996langevin}, is not closed.
Instead, it is determined by an infinite hierarchy involving the local harmonic densities whose $n$th-orders are defined by
\begin{equation}
    c_{\bm{r}}^{(n)} = \sum_{j=1}^N \cos(n\theta_j)\delta(\bm{r}-\bm{r}_j),
\end{equation}
and
\begin{equation}
    s_{\bm{r}}^{(n)} = \sum_{j=1}^N \sin(n\theta_j)\delta(\bm{r}-\bm{r}_j).
\end{equation}
The particle density corresponds to the 0th order harmonic density $\rho_{\bm{r}}=c_{\bm{r}}^{(0)}$, and the first-order harmonic densities are distinguished as the local polarization density $\bm{p}_{\bm{r}} = (c_{\bm{r}}^{(1)},s_{\bm{r}}^{(1)}) = \sum_{j=1}^N \bm{e}(\theta_j) \delta(\bm{r}-\bm{r}_j)$.
However, it was shown out that the infinite hierarchy of equations can be truncated at the first order in the hydrodynamics limit~\cite{cates2013active}.
In what follows, we shall ignore the higher-order harmonics.

The truncated coupled Dean equations are then given by
\begin{widetext}
\begin{equation}\label{eq:Dean_eq_0th}
    \dot{\rho}_{\bm{r}} 
    = -\bm{\nabla}_{\bm{r}} \cdot \left( v_0 \bm{p}_{\bm{r}}
    - \mu \int_V d\bm{r}'~ \rho_{\bm{r}} \rho_{\bm{r}'} \bm{\nabla}_{\bm{r}} \phi(|\bm{r}-\bm{r}'|)
    - D_t \bm{\nabla}_{\bm{r}} \rho_{\bm{r}}
    + \bm{\zeta}_{\bm{r}}^{\rho} \right),
\end{equation}
\begin{equation}\label{eq:Dean_eq_polar}
    \dot{\bm{p}}_{\bm{r}}
    = -\frac{v_0}{2} \bm{\nabla}_{\bm{r}} \rho_{\bm{r}}
    + \bm{\nabla}_{\bm{r}} \cdot \left( \mu \int_V d\bm{r}' ~
    \bm{p}_{\bm{r}} \rho_{\bm{r}'} \bm{\nabla}_{\bm{r}} \phi(|\bm{r}-\bm{r}'|)
    + D_t \bm{\nabla}_{\bm{r}} \bm{p}_{\bm{r}}
    - \bm{\zeta}_{\bm{r}}^{\bm{p},{\rm tr}}
    \right)
    - D_r \bm{p}_{\bm{r}}
    - \zeta_{\bm{r}}^{\bm{p},{\rm rot}},
\end{equation}
\end{widetext}
where the noise fields are constructed as
\begin{equation}\label{eq:noise_def1}
    \bm{\zeta}_{\bm{r}}^{\rho} 
    = \sqrt{2D_t} \sum_{j=1}^N \bm{\xi}_j \delta(\bm{r}-\bm{r}_j),
\end{equation}
\begin{equation}
    \bm{\zeta}_{\bm{r}}^{\bm{p},{\rm tr}}
    = \sqrt{2D_t} \sum_{j=1}^N \bm{\xi}_j 
    \begin{pmatrix}
        \cos \theta_j \\ \sin \theta_j
    \end{pmatrix} \delta(\bm{r}-\bm{r}_j),
\end{equation}
\begin{equation}\label{eq:noise_def3}
    \zeta_{\bm{r}}^{\bm{p},{\rm rot}} 
    = \sqrt{2D_r} \sum_{j=1}^N \eta_j
    \begin{pmatrix}
        \cos \theta_j \\ \sin \theta_j
    \end{pmatrix} \delta(\bm{r}-\bm{r}_j).
\end{equation}

The nonlinear terms in \eqref{eq:Dean_eq_0th} and \eqref{eq:Dean_eq_polar} can be linearized by assuming the fluctuations around the homogeneous state are small.
Since the homogeneous state does not exhibit any spatial discrete symmetry, the homogenous densities are  $\bar{\rho}=N/V$ and $\bar{\bm{p}} = 0$.
The fluctuations around this homogenous state are then denoted by $\rho_{\bm{r}} = \bar{\rho} + \delta\rho_{\bm{r}}$ and $\bm{p}_{\bm{r}} = \delta \bm{p}_{\bm{r}}$.
Neglecting the second-order terms in the fluctuations $\delta\rho_{\bm{r}} \delta\rho_{\bm{r}'}$ and $\delta \bm{p}_{\bm{r}} \delta\rho_{\bm{r}'}$, we can approximate the interaction as
\begin{equation}\label{eq:linearization}
\begin{aligned}
    & \int_V d\bm{r}' ~ X_{\bm{r}} \rho_{\bm{r}'} \bm{\nabla}_{\bm{r}}\phi(|\bm{r}-\bm{r}'|) \\
    & \approx \bar{X} \bm{\nabla}_{\bm{r}}
    \int_V d\bm{r}' ~ \delta\rho_{\bm{r}'} \phi(|\bm{r}-\bm{r}'|)
\end{aligned}
\end{equation}
for $X \in \{ \rho,~\bm{p} \}$.
Applying \eqref{eq:linearization} to \eqref{eq:Dean_eq_0th} and \eqref{eq:Dean_eq_polar}, and taking the Fourier transform in space, we obtain the following linearized equations:
\begin{equation}\label{eq:lin_Dean_eq_0th}
    \delta\dot{\rho}_{\bm{k}} 
    = i\bm{k} \cdot v_0 \bm{p}_{\bm{k}}
    - k^2 \left( D_t + \mu \bar{\rho} \phi_k \right) \delta\rho_{\bm{k}}
    + i\bm{k} \cdot \bm{\zeta}_{\bm{k}}^{\rho},
\end{equation}
\begin{equation}\label{eq:lin_Dean_eq_polar}
    \dot{\bm{p}}_{\bm{k}}
    = \frac{1}{2} i\bm{k} v_0 \rho_{\bm{k}}
    - \left( D_r + k^2 D_t \right) \bm{p}_{\bm{k}}
    + i\bm{k} \cdot \bm{\zeta}_{\bm{k}}^{\bm{p},{\rm tr}}
    - \zeta_{\bm{k}}^{\bm{p},{\rm rot}},
\end{equation}
where $\phi_k = \int d\bm{r} e^{i\bm{k}\cdot{\bm{r}}} \phi(r)$.
This type of linearization of the Dean equation is known to be accurate for soft interactions at a high density~\cite{demery2014generalized}.

Since all the steady-state covariances of the noise fields in (\ref{eq:noise_def1}-\ref{eq:noise_def3}) are constant, we can replace multiplicative noises in \eqref{eq:lin_Dean_eq_0th} and \eqref{eq:lin_Dean_eq_polar} with additive noises in the homogeneous steady-state.
Consequently, in the homogeneous steady state, \eqref{eq:lin_Dean_eq_0th} and \eqref{eq:lin_Dean_eq_polar} become coupled linear stochastic equations with additive noises:
\begin{equation}\label{eq:lin_Dean_eq_final}
    \begin{pmatrix}
        \delta\dot{\rho}_{\bm{k}} \\
        i\bm{k} \cdot \dot{\bm{p}}_{\bm{k}}
    \end{pmatrix}
    = -\mathsf{L} \cdot 
    \begin{pmatrix}
        \delta\rho_{\bm{k}} \\
        i\bm{k} \cdot \bm{p}_{\bm{k}}
    \end{pmatrix}
    + \begin{pmatrix}
        \sqrt{2k^2 D_t N} \Lambda_{\bm{k}}^{(0)} \\
        \sqrt{k^2 D_r N} \Lambda_{\bm{k}}^{(1)}
    \end{pmatrix}
\end{equation}
with matrix
\begin{equation}
    \mathsf{L} = \begin{pmatrix}
        k^2 (D_t + \mu\bar{\rho} \phi_0) & -v_0 \\
        k^2 v_0/2 & D_r
    \end{pmatrix}.
\end{equation}
The noise fields $\Lambda_{\bm{k}}^{(n)}(t)$ are characterized by $\langle \Lambda_{\bm{k}}^{(m)}(t) \Lambda_{\bm{k}'}^{(n)}(t')\rangle_{\rm ss} = \delta_{mn} \delta(\bm{k}+\bm{k}')\delta(t-t')$.

Equation~\eqref{eq:lin_Dean_eq_final} belongs to the class of the multivariate Ornstein-Uhlenbeck processes, whose covariance matrix is analytically solvable.
We define the steady-state covariance matrix of $\delta\rho_{\bm{k}}$ and $i\bm{k}\cdot\bm{p}_{\bm{k}}$ as \begin{equation}
    \mathsf{\Sigma} 
    = \begin{pmatrix}
        \langle \delta \rho_{\bm{k}} \delta \rho_{-\bm{k}} \rangle_{\rm ss}
        & \langle \delta \rho_{\bm{k}} (-i\bm{k} \cdot \bm{p}_{-\bm{k}}) \rangle_{\rm ss} \\
        \langle (i\bm{k} \cdot \bm{p}_{\bm{k}}) \delta\rho_{-\bm{k}} \rangle_{\rm ss}
        & \langle (i\bm{k} \cdot \bm{p}_{\bm{k}}) (-i\bm{k} \cdot \bm{p}_{-\bm{k}}) \rangle_{\rm ss}
    \end{pmatrix},
\end{equation}
which satisfies the matrix algebraic equation $\mathsf{L} \cdot \mathsf{\Sigma} + \mathsf{\Sigma} \cdot \mathsf{L}^{\rm T} = k^2 N {\rm diag}\{2D_t, D_r \}$~\cite{gardiner2009stochastic}.
The two-time correlation function of $\delta\rho_{\bm{k}}$ is given by $\langle \delta\rho_{\bm{k}}(t) \delta\rho_{-\bm{k}}(0) \rangle_{\rm ss} = [e^{-t\mathsf{L}} \cdot \mathsf{\Sigma}]_{11}$.
Thus we have
\begin{equation}
\begin{aligned}
    \tilde{\chi}
    & = \lim_{k\to 0} \langle \delta\rho_{\bm{k}} \delta\rho_{-\bm{k}} \rangle_{\rm ss} \\
    & = \lim_{k\to 0} [\mathsf{\Sigma}]_{11}
    = \frac{D_t + v_0^2/(2D_r)}{D_t + \mu \bar{\rho} \phi_0 + v_0^2/(2D_r)} N,
\end{aligned}
\end{equation}
\begin{equation}
    E
    = -\lim_{k\to 0} \frac{\langle\delta\dot{\rho}_{\bm{k}} \delta\rho_{-\bm{k}} \rangle_{\rm ss}}{k^2}
    = \lim_{k\to 0} \frac{[\mathsf{L} \cdot \mathsf{\Sigma}]_{11}}{k^2}
    =D_t N,
\end{equation}
\begin{equation}
\begin{aligned}
    C 
    & = \lim_{\omega \to 0} \lim_{k\to 0} \frac{\langle\delta\dot{\rho}_{\bm{k}\omega} \delta\dot{\rho}_{-\bm{k}} \rangle_{\rm ss}}{k^2} \\
    & = -\lim_{\omega \to 0}\lim_{k\to 0} \frac{[\mathsf{L}^2 \cdot (-i\omega \mathsf{I} + \mathsf{L})^{-1} \cdot \mathsf{\Sigma}]_{11}}{k^2} \\
    & = \frac{v_0^2}{2D_r} N.
\end{aligned}
\end{equation}
In conclusion, the linearized Dean equation predicts that the transport coefficient is given by
\begin{equation}
    D = \tilde{\chi}^{-1} ( C + E )
    = D_t + \mu\bar{\rho}\phi_0 + \frac{v_0^2}{2D_r}.
\end{equation}
Noteworthy is that $\delta\rho_{\bm{k}}$ is the same as $\rho_{\bm{k}}$ as long as $\bm{k}$ is nonzero since the offset $\int_V d\bm{r} ~ e^{i\bm{k}\cdot\bm{r}} \bar{\rho} = N \delta(\bm{k})$ only contributes at $\bm{k}=0$.

\subsection{Numerical simulations}
We measure the transport coefficient $D$ in two independent numerical simulations.
ABP simulations were performed using bespoke computer code implementing the Euler algorithm in Python v2.7.16.
The length scale was set by $a=1$, the force scale by setting $\mu=1$, and the time scale by setting $D_r=10$ and $D_t=0$.
All simulations were performed at a density of ${\bar \rho}=1$.
The integration time step are chosen to be $\Delta t = 0.001$ for relaxation simulations and $\Delta t = 0.005$ for steady-state simulations.
The simulations are performed for 10 combinations of two interaction strengths $K \in \{0.5,1 \}$ and five self-propulsion speeds $v_0 \in \{ 1,2,3,4,5 \}$.

\subsubsection{relaxation simulation}
We first measure $D$ directly by observing the regression of $\rho_{\bm{k}}$.
Due to the noise in the data, we only use the smallest-$k$ mode to measure $D$.
Initially, we place $N = 1600$ particles uniformly in the region between $x=L/4$ and $x=3L/4$ and $y$ from $0$ to $L$, with $L=40$, and then simulate the particle dynamics for $10^5$ time steps.
The time series of the Fourier modes of the local particle number density is calculated from the position data. 
We only observe the $x$-directional mode, i.e., $\bm{k} = (2\pi/L, 0)$.
For a given parameter set $(K,v_0)$, we perform $10$ independent simulations to get 10 time series of $\rho_{\bm{k}}$.
Each mode $\rho_{\bm{k}}$ decays exponentially with an exponent $-1/\tau$.
We first measure the decay exponent $1/\tau$ for 10 samples, and then estimate $D$ from $\tau/(2\pi/L)^2$.
Taking the average and standard error on the mean over 10 independently measured $D$, we estimate the transport coefficient and its error.

\subsubsection{steady-state simulation}
Second, we deduce the transport coefficient from the Green-Kubo relation $D\tilde{\chi} = C + E$ where
\begin{equation}\label{eq:GK_ABPs_simulation}
\begin{aligned}
    \tilde{\chi} 
    & = \lim_{k\to 0} \langle \rho_{\bm{k}} \rho_{-\bm{k}} \rangle_{\rm ss}, \\
    C 
    & = \lim_{k\to 0} \int_0^\infty dt ~ \langle j_{\bm{k}}(t) j_{-\bm{k}}(0) \rangle_{\rm ss}, \\
    E
    & = \lim_{k\to 0} 
    \frac{\langle j_{\bm{k}}\rho_{-\bm{k}} \rangle_{\rm ss}}{ik}.
\end{aligned}
\end{equation}
Since the particles do not prefer any particular direction, the drift coefficient vanishes $v=0$.
To this end, we measure the steady-state correlation functions $\langle \rho_{\bm{k}} \rho_{-\bm{k}} \rangle_{\rm ss}$, $\langle j_{\bm{k}} \rho_{-\bm{k}} \rangle_{\rm ss}$, and $\langle j_{\bm{k}}(t) j_{-\bm{k}}(0) \rangle_{\rm ss}$ numerically.
The local current $j_{\bm{k}}$ is calculated by taking the $x$-component of
\begin{equation}\label{eq:current_simulation}
\begin{aligned}
    \bm{j}_{\bm{k}}(t)
    =\sum_{j=1}^N \left( \frac{e^{i\bm{k} \cdot \bm{r}_j(t)} + e^{i\bm{k} \cdot \bm{r}_j(t-\Delta t)}}{2} \right) \\
    \times \left( \frac{\bm{r}_j(t)-\bm{r}_j(t-\Delta t)}{\Delta t} \right).
\end{aligned}
\end{equation}
The system size is chosen to be $L = 20$ for the steady-state simulations.
To estimate the limiting value of the correlation functions for $k\to 0$, we use the smallest-$k$ mode as an approximation instead of taking the extrapolation from several modes.
This is because the extrapolation is not reliable due to the noise in data and the small system size.
To calculate the steady-state correlation functions, we assume that the system is ergodic and replace the ensemble average with a time average, i.e., $\langle O_1(t) O_2(s) \rangle_{\rm ss} = \lim_{\mathcal{T} \to \infty} \frac{1}{\mathcal{T}} \int_0^\mathcal{T} dt' O_1(t+t') O_2(s+t')$ for any observables $O_1$ and $O_2$.
We run a single long simulation for $4\times 10^6$ time steps and drop the first $8\times 10^5$ steps to discard transient dynamics.

The details of estimating the three quantities in~\eqref{eq:GK_ABPs_simulation} are as follows.
First, we estimate $\tilde{\chi}$ as the correlation function $\langle \rho_{\bm{k}} \rho_{-\bm{k}} \rangle_{\rm ss}$ of the smallest-$k$ mode.
Second, we estimate $E$ by the slope of ${\rm Im}\{\langle j_{\bm{k}} \rho_{-\bm{k}} \rangle \}$ as a linear function of $k$ in the small-$k$ regime.
To measure the slope, we use the first two smallest-$k$ modes.
Lastly, assuming exponential decay of $\langle j_{\bm{k}}(t) j_{-\bm{k}}(0) \rangle_{\rm ss} = \langle j_{\bm{k}} j_{-\bm{k}} \rangle_{\rm ss} e^{-t/\tau}$, we estimate $C$ from $\tau \langle j_{0}^2 \rangle_{\rm ss}$. 
The two-time correlation $\langle j_0(t) j_0(0) \rangle_{\rm ss}$ is calculated using the Wiener-Khinchin theorem~\cite{gardiner2009stochastic}, and $\tau$ is estimated from a least-square linear fitting on $\ln \langle j_0(t) j_0(0) \rangle_{\rm ss}$ as a function of $t$.
To analyze the error, we divide the time series of $\rho_{\bm{k}}$ and $j_{\bm{k}}$ into 10 blocks and repeat the estimation of $D$ for each block of data.
The error of $D$ is estimated by the standard error on the mean over the 10 estimations.

\section{Noisy Kuramoto model}\label{sec:Kuramoto}
\subsection{Linear approximation}
In this section, we derive a linear approximation to the noisy Kuramoto model near the synchronization state.
Thanks to the linearity, correlation functions can be obtained analytically, and thus the transport coefficient of the Nambu-Goldstone mode can be deduced from a Green-Kubo relation.

We recall the stochastic equation that governs the time evolution of the $i$-th oscillator~\eqref{eq:Kuramoto_model}:
\begin{equation}\label{eq:Kuramoto_Appdx}
    \dot{\theta}_i(t)
    = \Omega_i + K \sum_{j\sim i}^N \sin(\theta_j(t) - \theta_i(t)) + \sqrt{2T}\xi_i(t)
\end{equation}
where the sum extends over neighbors of $i$.
The oscillators are evenly spaced on a two-dimensional square lattice and the position of the $i$-th particle is denoted by $\bm{r}_i$.
The intrinsic frequency $\Omega_i$ is sampled from a Gaussian distribution with mean  $\bar{\Omega}$ and variance $\sigma^2$.
In the synchronized state, the oscillators rotate coherently at a common frequency $\bar{\Omega}$, so that each oscillator deviates slightly $\theta_i(t) = \theta_0 + \bar{\Omega}t + \delta\theta_i(t)$, where $\theta_0$ is an arbitrary offset.
Assuming the phase differences between neighboring oscillators are small, we approximate the interaction in~\eqref{eq:Kuramoto_Appdx} as
\begin{equation}
\begin{aligned}
    K\sum_{j\sim i} \sin(\theta_j - \theta_i) 
    & \approx K \sum_{j\sim i} (\theta_j - \theta_i) \\
    & = K \sum_{j\sim i} (\delta\theta_j - \delta\theta_i).
\end{aligned}
\end{equation}
Then near the synchronized state, the time evolution of the Fourier modes $\delta\theta_{\bm{k}} = \sum_j \delta\theta_j e^{i\bm{k}\cdot\bm{r}_j}$ is governed by the equation
\begin{equation}\label{eq:Kuramoto_apprx}
\begin{aligned}
    \dot{\delta\theta}_{\bm{k}}
    & = \Omega_{\bm{k}} - 2K \left( 2 - \cos k_x - \cos k_y \right) \delta\theta_{\bm{k}}(t) \\
    & ~~~ + \sqrt{2TN}\zeta_{\bm{k}}(t),
\end{aligned}
\end{equation}
with noise correlations $\langle \zeta_{\bm{k}}(t) \zeta_{\bm{k}'}(t') \rangle = \delta(\bm{k} + \bm{k}') \delta(t-t')$.
For long-wavelength modes, we can linearize \eqref{eq:Kuramoto_apprx} in $k$ to obtain the approximated linear equation
\begin{equation}
    \dot{\delta\theta}_{\bm{k}}(t)
    = \Omega_{\bm{k}} - k^2 K \delta\theta_{\bm{k}}(t) + \sqrt{2TN}\zeta_{\bm{k}}(t),
\end{equation}
with solution
\begin{equation}
\begin{aligned}
    \delta\theta_{\bm{k}}(t) 
    & = e^{-tk^2K} \delta\theta_{\bm{k}}(0) \\
    & + \int_0^t dt' ~ e^{-(t-t')k^2 K} \left( \Omega_{\bm{k}} + \sqrt{2TN} \zeta_{\bm{k}}(t') \right).
\end{aligned}
\end{equation}
Utilizing this solution, we can obtain the steady-state equal-time correlation function from the long time
limit
\begin{equation}
\begin{aligned}
    & \langle \delta\theta_{\bm{k}} \delta\theta_{-\bm{k}} \rangle_{\rm ss} \\
    & = \lim_{t\to\infty} \langle \delta\theta_{\bm{k}}(t) \delta\theta_{-\bm{k}}(t) \rangle
    = \frac{\langle \Omega_{\bm{k}}\Omega_{-\bm{k}} \rangle }{k^4K^2} + \frac{TN}{k^2 K},
\end{aligned}
\end{equation}
which depends on the disorder.
Since the intrinsic frequency of each oscillator is independent of the others, the frequency correlation function is given by
\begin{equation}
\begin{aligned}
    \langle \Omega_{\bm{k}} \Omega_{-\bm{k}} \rangle
    & = \sum_{j=1}^N \sum_{l=1}^N \langle \Omega_j \Omega_{l} \rangle e^{i\bm{k}\cdot (\bm{r}_j - \bm{r}_l)} \\
    & = \sum_{j=1}^N \sum_{l=1}^N \sigma^2 \delta_{jl}  e^{i\bm{k}\cdot (\bm{r}_j - \bm{r}_l)}
    = \sigma^2 N,
\end{aligned}
\end{equation}
leading to
\begin{equation}\label{eq:corr1_Kuramoto}
    \langle \delta\theta_{\bm{k}} \delta\theta_{-\bm{k}} \rangle_{\rm ss} 
    = \frac{\sigma^2 N}{k^4 K^2}\left( 1 + \frac{K T}{\sigma^2} k^2 \right).
\end{equation}
Other correlation functions involving time-derivatives can be obtained from the two-time correaltion function
\begin{equation}
\begin{aligned}
    \langle \delta\theta_{\bm{k}}(t) \delta\theta_{-\bm{k}}(0) \rangle_{\rm ss}
    & = \langle \delta\theta_{\bm{k}} \delta\theta_{-\bm{k}} \rangle_{\rm ss} e^{-t k^2 K} \\
    & = \frac{\sigma^2 N}{k^4 K^2}\left( 1 + \frac{K T}{\sigma^2} k^2 \right) e^{-t k^2 K}.
\end{aligned}
\end{equation}
Explicitly, we have
\begin{equation}\label{eq:corr2_Kuramoto}
\begin{aligned}
    \langle \dot{\delta\theta}_{\bm{k}} \delta\theta_{-\bm{k}} \rangle_{\rm ss}
    & = \lim_{t\to 0} \partial_t \langle \delta\theta_{\bm{k}}(t) \delta\theta_{-\bm{k}}(0) \rangle_{\rm ss} \\
    & = -\frac{\sigma^2 N}{k^2 K}\left( 1 + \frac{K T}{\sigma^2} k^2 \right),
\end{aligned}
\end{equation}
\begin{equation}
\begin{aligned}
    \langle \dot{\delta\theta}_{\bm{k}\omega} \dot{\delta\theta}_{-\bm{k}} \rangle_{\rm ss}
    & = -\int_0^\infty dt ~ e^{i\omega t} \partial_t^2 \langle \delta\theta_{\bm{k}}(t) \delta\theta_{-\bm{k}}(0) \rangle_{\rm ss} \\
    & = \frac{\sigma^2 N}{i\omega - k^2 K}\left( 1 + \frac{K T}{\sigma^2} k^2 \right).
\end{aligned}
\end{equation}
Green-Kubo relations~\eqref{eq:GK_v_NG} and~\eqref{eq:GK_D_NG} give the transport coefficients as
\begin{equation}
    v = \lim_{\omega \to 0}\lim_{k\to 0} \frac{\langle \dot{\delta\theta}_{\bm{k}} \delta\theta_{-\bm{k}} \rangle_{\rm ss}
    - \langle \dot{\delta\theta}_{\bm{k}\omega} \dot{\delta\theta}_{-\bm{k}} \rangle_{\rm ss}}{ik \langle \delta\theta_{\bm{k}} \delta\theta_{-\bm{k}} \rangle_{\rm ss}}
    = 0,
\end{equation}
and
\begin{equation}
    D = \lim_{\omega \to 0}\lim_{k\to 0} \frac{\langle \dot{\delta\theta}_{\bm{k}\omega} \dot{\delta\theta}_{-\bm{k}} \rangle_{\rm ss}
    - \langle \dot{\delta\theta}_{\bm{k}} \delta\theta_{-\bm{k}} \rangle_{\rm ss}
    }{k^2 \langle \delta\theta_{\bm{k}} \delta\theta_{-\bm{k}} \rangle_{\rm ss}}
    = K.
\end{equation}

\subsection{Numerical simulations}
We measure the transport coefficient $D$ in two independent numerical simulations.
Simulations of the noisy Kuramoto model were performed using bespoke computer code implementing the Heun algorithm~\cite{greiner1988numerical} in Python v3.7.4.
The integration time step are chosen to be $\Delta t = 0.001$ for the relaxation simulations and $\Delta t = 0.01$ for the steady-state simulations.
The phases of oscillators are defined in the range of [$-\pi$,$\pi$], and the average phase is set to zero.
This is equivalent to choosing a co-moving frame of reference.
The simulations are performed for 8 combinations of four interaction strengths $K \in \{1,~2,~3,~4 \}$ and two noise strengths $(T,\sigma) \in \{ (0.002,0.005),(0.01,0.01) \}$.

\begin{figure}\centering
	\includegraphics[width=\columnwidth]{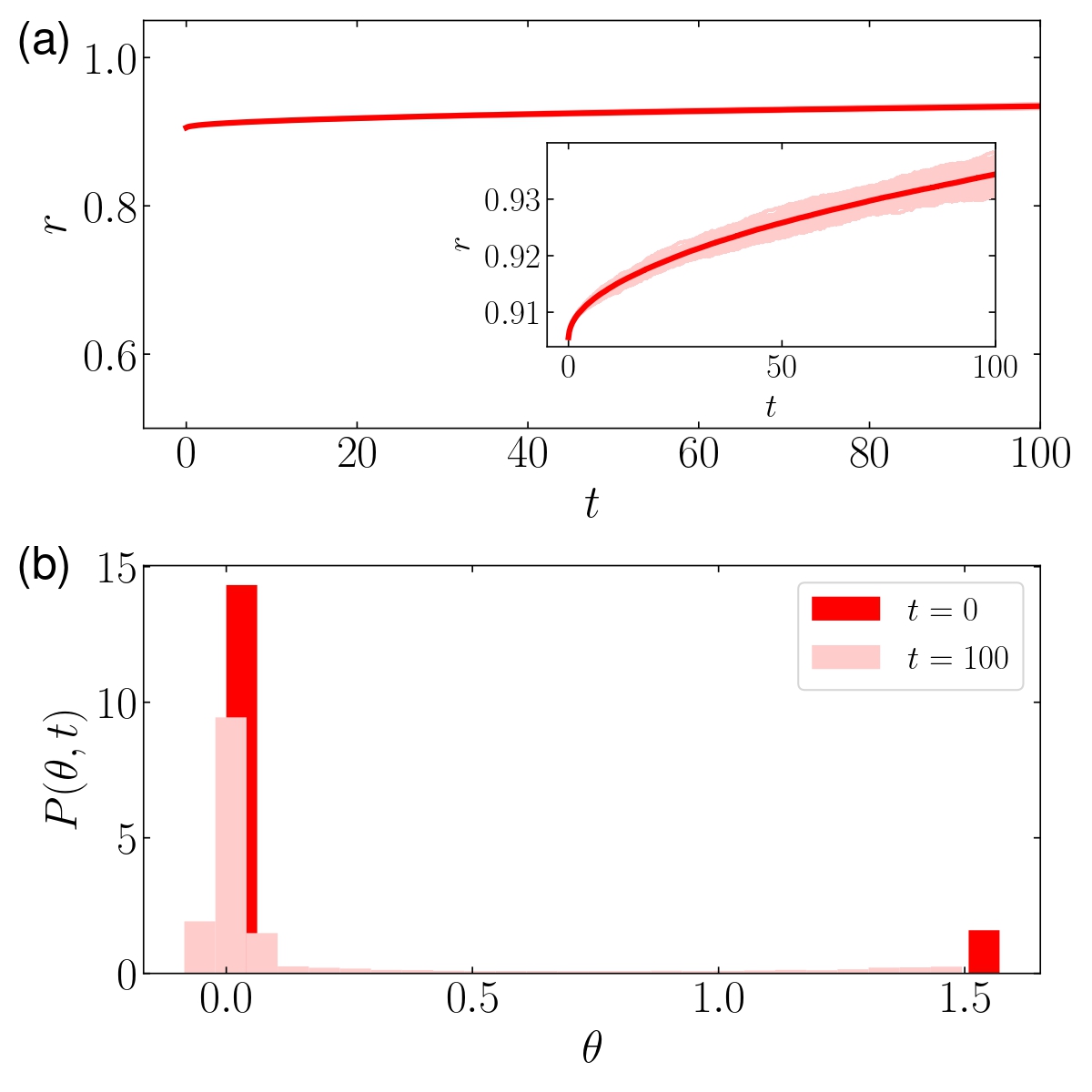}
	\caption{The relaxation to the synchronized steady state is observed from the changes of two quantities.
	(a) The order parameter increases with time.
	The pink lines show the changes in the order parameter for 100 independent simulations.
	The red line shows the change in the ensemble-averaged order parameter, $|\sum_{j=1}^N \langle e^{i\theta_j} \rangle_{\rm ss}|$.
	(b) The initial bimodal phase distribution approaches a unimodal distribution as time passes.
	The system parameters are chosen as $(K, T, \sigma) = (4,0.002,0.005)$.}
	\label{fig:Kuramoto_relaxation}
\end{figure}

\subsubsection{relaxation simulation}
We first measure $D$ directly by observing the regression of $|\langle \theta_{\bm{k}} \rangle|$ for the first five modes with the smallest $k$ values.
To obtain the average $\langle \theta_{\bm{k}} \rangle$, 100 independent simulations are performed for a given parameter set $(K,T,\sigma)$.
The intrinsic frequencies $\Omega_i$ and the noise trajectories $\xi_i(t)$ vary by sample.
The initial phases are assigned to be $\pi/2$ for a leftmost tenth of the oscillators and $0$ for the others.
The system size is $L_x = 1000$ and $L_y = 20$ and simulations run for $10^5$ steps.
The time series of $\theta_{\bm{k}}$ is calculated from the phase data. 
Only the $x$-direction modes are observed so as to see the slowest regressions.
That is, the wave vectors of the observed modes are $\bm{k} = (2\pi/L_x) \bm{n}$ with $ \bm{n} \in \{(1,0),~(2,0),~(3,0),~(4,0),~(5,0)\}$.
The averaged modes $|\langle \theta_{\bm{k}} \rangle|$ decay exponentially with exponent $-1/\tau(k)$.
We first measure $1/\tau(k)$ for the five different $k$ by least-square linear fittings of $\ln|\langle \theta_{\bm{k}} \rangle|$.
Then a quadratic fitting of $1/\tau(k)$ provides an estimate for $D$.
The error is estimated from the standard deviation of the quadratic fitting.

The relaxation from the inhomogeneous initial state to the synchronized homogeneous steady state is also observed from two other quantities.
The order parameter of the Kuramoto model, defined by the absolute value of $r e^{i\varphi} = \frac{1}{N} \sum_{j=1}^N e^{i\theta_j}$, quantifies the extent to which the phases of oscillators are synchronized.
FIG.~\ref{fig:Kuramoto_relaxation}(a) shows the order parameter of the system increases with time, indicating a relaxation to the synchronized steady-state.
Also, the phase distribution $P(\theta,t)$ in FIG.~\ref{fig:Kuramoto_relaxation}(b) shows the unsynchronized phases ($\theta = \pi/2 $) gradually absorb into around the synchronized phases ($\theta = 0$).

\subsubsection{steady-state simulation}
Second, we deduce the transport coefficient from the Green-Kubo relation~\eqref{eq:GK_NG_modes}.
Since the linear theory predicts $\langle \dot{\theta}_{\bm{k}\omega} \dot{\theta}_{-\bm{k}} \rangle_{\rm ss}$ is much smaller than $\langle \dot{\theta}_{\bm{k}} \theta_{-\bm{k}} \rangle_{\rm ss}$ by the factor $k^2$, we estimate the transport coefficient by the ratio
\begin{equation}
    D = -\lim_{k\to 0} \frac{\langle \dot{\theta}_{\bm{k}} \theta_{-\bm{k}} \rangle_{\rm ss}}{k^2 \langle \theta_{\bm{k}} \theta_{-\bm{k}} \rangle_{\rm ss}}.
\end{equation}
To this end, we measure the steady-state correlation functions $\langle \theta_{\bm{k}} \theta_{-\bm{k}} \rangle_{\rm ss}$ and $\langle \dot{\theta}_{\bm{k}} \theta_{-\bm{k}} \rangle_{\rm ss}$ numerically.
The system size is chosen by $L_x = L_y = L = 20$ for the steady-state simulations.
The first 18 modes with the smallest $k$ values are observed.
That is, the wave vectors of the observed modes are $\bm{k} = (2\pi/L) \bm{n}$ with $|\bm{n}| \in \{ 1, \sqrt{2}, 2, \sqrt{5}, 3,~\sqrt{8},~\sqrt{10}\}$.
The degeneracy of the modes are $\{ 2, 2, 2, 4, 2, 2, 4 \}$, respectively.
To calculate the steady-state correlation functions, we assume that the system is ergodic and replace the ensemble average with a time average, i.e., $\langle O_1(t) O_2(s) \rangle_{\rm ss} = \lim_{\mathcal{T} \to \infty} \frac{1}{\mathcal{T}} \int_0^\mathcal{T} dt' O_1(t+t') O_2(s+t')$ for any observables $O_1$ and $O_2$.
Since the linear theory predicts the relaxation time is $\tau(k) = 1/(k^2 K)$, we run a single long simulation of ($5\times 10^6$ + $\tau(k=2\pi/L_x)/\Delta t$) time steps and drop the first $\tau(k=2\pi/L_x)/\Delta t$ steps to discard transient dynamics.
We perform a weighted quadratic fitting on $\langle \dot{\theta}_{\bm{k}} \theta_{-\bm{k}} \rangle_{\rm ss}/(k^2 \langle \theta_{\bm{k}} \theta_{-\bm{k}} \rangle_{\rm ss})$ with respect to $k$ so as to get its limiting value for $k\to 0$.
The weighting factor is the inverse square of the error.
We estimate $D$ and its error by the fitting value $a$ of the fitting function $a + bk^2$ and its standard deviation, respectively.

\begin{figure}
	\includegraphics[width=\columnwidth]{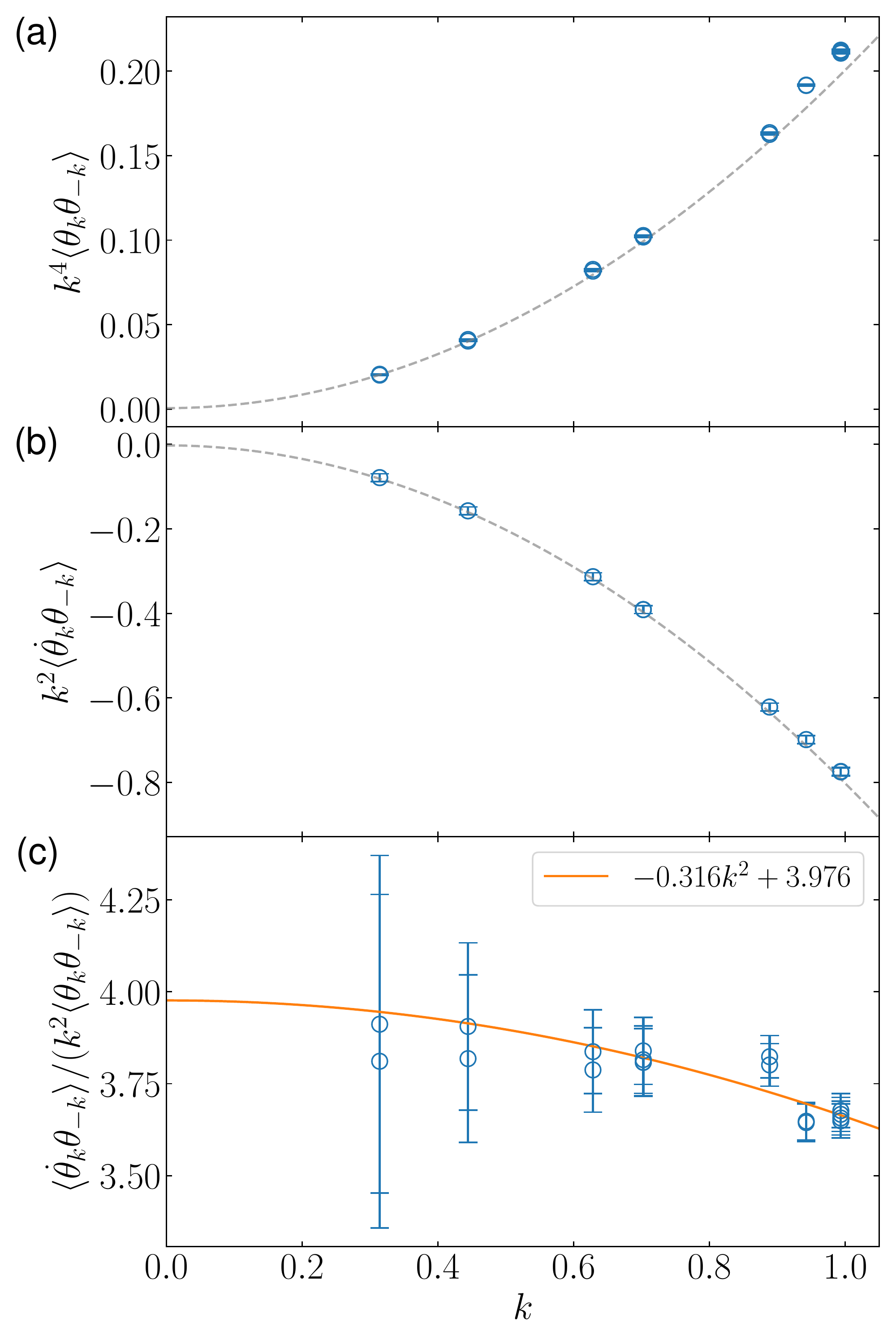}
	\caption{Steady-state correlation functions of the noisy Kuramoto model for a set of parameters $(K,T,\sigma) = (4,0.002,0.005)$.
	The dashed lines in (a) and (b) are the correlation functions obtained from the linear approximation,~\eqref{eq:corr1_Kuramoto} and~\eqref{eq:corr2_Kuramoto}.
	The orange solid line in (c) is obtained by a weighted least-square fitting, and the legend shows the fitting values.}
	\label{fig:Kuramoto_steady_state}
\end{figure}

To estimate errors of correlation functions, we found the effective independent number of data points $n_{\rm eff} = t_{\rm run}/(2t_{\rm corr}(k))$ by measuring the correlation times of $\theta_{\bm{k}}(t) \theta_{-\bm{k}}(t)$ and $
\dot{\theta}_{\bm{k}}(t) \theta_{-\bm{k}}(t)$ directly, where $t_{\rm run} = 5\times 10^3$ is the simulation time and $t_{\rm corr}(k)$ is numerically measured correlation time.
The errors are estimated by the standard deviations of $\theta_{\bm{k}}(t) \theta_{-\bm{k}}(t)$ and $
\dot{\theta}_{\bm{k}}(t) \theta_{-\bm{k}}(t)$ divided by the factor $\sqrt{n_{\rm eff}}$~\cite{allen1989computer}.
The error of the ratio between two correlation functions is determined by the rule of propagation of errors.
FIG.~\ref{fig:Kuramoto_steady_state} exemplifies the $k$-dependence of steady-state correlation functions for a fixed set of parameters.

\bibliographystyle{apsrev}
\bibliography{paper.bib}

\end{document}